\documentclass[twocolumn,amssymb,fleqn]{revtex4} 
\usepackage{epsfig,amssymb,amsmath,graphicx,subfigure,hyperref  }  
\usepackage{color,slashbox}

\newcommand{\be}{\begin{eqnarray}}   
\newcommand{\ee}{\end{eqnarray}}

\begin{document}

\title{Electrostatic repulsion-driven crystallization model arising from filament networks}
\author{Zhenwei Yao and Monica Olvera de la Cruz}
\affiliation{Department of Materials Science and Engineering,
Northwestern University, Evanston, Illinois 60208-3108, USA}
\begin{abstract}
The crystallization of bundles in filament networks interacting via long-range
repulsions in confinement is described by a phenomenological model. The model
demonstrates the formation of the hexagonal crystalline order via the interplay
of the confinement potential and the filament-filament repulsion. Two distinct
crystallization mechanisms in the short- and large- screening length regimes are
discussed, and the phase diagram is obtained. Simulation of large bundles
predicts the existence of topological defects within the bundled filaments. This
electrostatic repulsion-driven crystallization model arising from studying
filament networks can even find a more general context extending to charged colloidal
systems.

\end{abstract}
 \maketitle


\section{Introduction}

Pattern formation from mutually repelling units in confined geometries has
inspired various experimental~\cite{bausch2003grain,irvine2010pleats} and
theoretical~\cite{levin2002electrostatic, solis2011electrostatic,
bowick2011crystalline, yao2013packing} studies. These patterns provide a route to directed
self-assembly~\cite{devries2007divalent}. Moreover, fascinating physics emerge
in confined geometries that  influence the physical properties of purely
repulsive particle systems. For example, topological defects in 2D crystalline
order on curved geometries, resulting from the repulsion of confining particles,
can influence the melting of 2D crystals~\cite{kosterlitz2002ordering} and the
mechanical properties of materials~\cite{nelson2002defects}.   Recent
experiments show the crystallization of like-charge synthetic supramolecular
peptide filaments into lattices with very large
spacings~\cite{cui2010spontaneous}. Bundles of crystallized filaments  are
observed to be randomly distributed forming a network of bundles. The
observed, unexpectedly large crystalline lattice spacing between crystallized
filaments in the bundle excludes the possibility of short-range attractions
associated with counterion correlations that occur between close rods or
filaments~\cite{wong2010electrostatics}.  The underlying crystallization
mechanism is therefore fundamentally distinct from those reported for
cytoskeleton filaments and ds-DNA strands in the presence of short-range
attractions, such as those induced by multi-valent counterions that lead to
the formation of compact bundles~\cite{widom1980cation,
raspaud1998precipitation,stendahl2006intermolecular,greenfield2009tunable,
PhysRevE.82.031901}.  Without any attractive interaction, the confinement effect
due to the observed network of bundles seems to be the only force to counter the
repulsion among filaments.  

In this work, we analyze the interplay between the
repulsive interaction and network confinement in the crystallization of
filaments.  We develop a particle model to understand how the long-range
repulsions induce hexagonal crystalline order inside bundles of filaments, where
the bundles form networks or gels. The electrostatic repulsion-driven
crystallization model arising from filament networks can even be discussed in a
more general context; in particular the introduced spatially varying confinement
potential can be employed to manipulate charged particles in general colloidal
systems~\cite{lowen2012colloidal}.

\section{Model}

The filaments in bundles are observed to be straight up to the scale of one
micron, while their cross sectional radius is only a few nanometers ~\cite{hartgerink2001self}; their deformation is neglected in our model.
By projecting these filaments to the plane perpendicular to them, the
three-dimensional problem of the disorder-order transition of bundled filaments
is reduced to the crystallization of particles in a confined flat disk; the
thickness of filaments are neglected given the large lattice spacing. In what
follows, we discuss the energetics of these particles. Experimental work
suggests that the Poisson-Boltzmann
equation provides a
reasonably accurate description of even highly charged polyelectrolytes in 1:1
solutions despite its mean field
nature, which is the case of interest in experiment~\cite{wong2010electrostatics}. The extraordinarily large
lattice spacing of crystallized filaments (in comparison with the screening length)
validates the application of the Debye-H$\ddot{\textrm{u}}$ckel solution to the PB
equation for the interaction energy between filaments; the possible counterion
correlations on polyelectrolyte surfaces are significantly diminished beyond a
very short distance~\cite{PhysRevLett.79.1289, diehl2001counterion}.
The screened Coulomb interaction energy between two parallel polyelectrolyte
cylinders is~\cite{brenner1974physical}: \begin{eqnarray}
  V_{int}(\vec{r}_i-\vec{r}_j) =  A K_0(\kappa ||\vec{r}_i-\vec{r}_j||)
  ,\label{VPB} \end{eqnarray} where $A$ is a constant related to charge
  densities on filaments, $K_0(x)$ is the zeroth-order modified Bessel
  function of the second kind, and $\kappa^{-1}$ is the Debye screening length.
  Note that the full cylindrical solution to the Poisson-Boltzmann equation
  includes terms of $Tr(K)=K_0(x)$ and $Tr(K^j)= {\cal O}(e^{-jx}) (j=3, 5, 7...)$
  as $x\rightarrow \infty$~\cite{tracy1997exact}. Therefore, the linearized approximation to the
  filament-filament interaction $V_{int}$ essentially neglects terms that decay
  faster than $e^{-\kappa r}$; the contributions from these terms are trivial for large
  distances between filaments~\cite{brenner1973theory, harries1998solving}. 
 For sufficiently large
  screening length, the interaction energy between particles takes the form of
  the two-dimensional Coulomb interaction that can be derived from the
  two-dimensional Poisson equation~\cite{mastatistical}:
  $V_{Coulomb}(\vec{r}_i-\vec{r}_j) =
  \frac{\lambda^2}{2\pi\epsilon}\ln(\frac{a}{||\vec{r}_i - \vec{r}_j||})$, where
  $\lambda$ is the line density of charges on filaments, and $a$ is a constant.
  Note that for $x\equiv \kappa r <<1$, $K_0(x) =-\ln(x)+ \ln 2-\gamma
  +{\cal{O}}(x^2)$, where the Euler constant $\gamma \approx 0.5772$. The
  neglect of the constant terms in the expansion for $K_0(x)$ also leads to the
  expression for the 2D Coulomb interaction. It is interesting to note that the
  interaction energy of two vortex lines in superconductors is also proportional
  to the zeroth-order modified Bessel function of the second kind as in Eq.(\ref{VPB})~\cite{tinkham2012introduction}.

\begin{figure}[t]  
\centering 
\subfigure[]{
\includegraphics[width=1.0in]{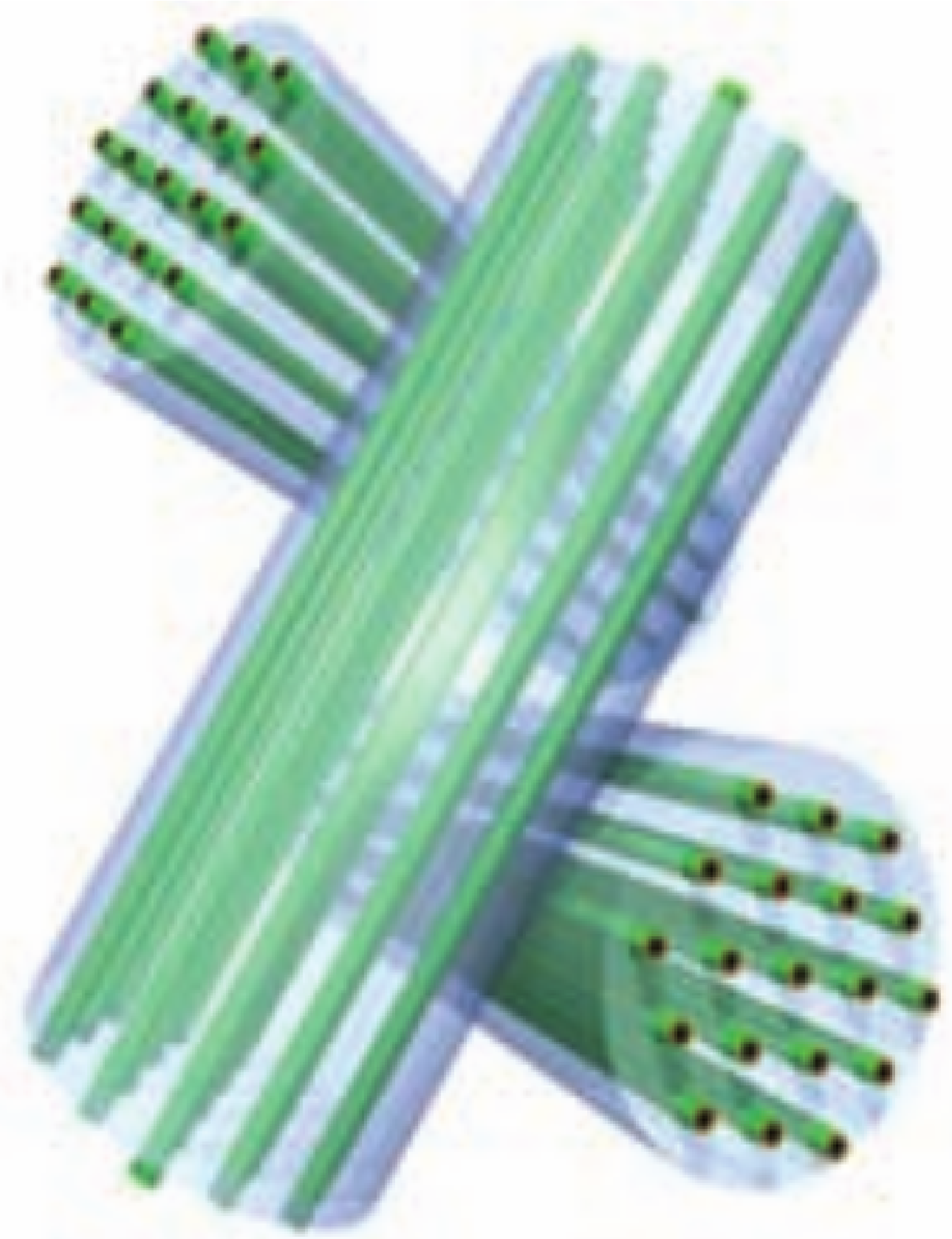}}
\hspace{0.2in} 
\subfigure[]{
\includegraphics[width=1.3in]{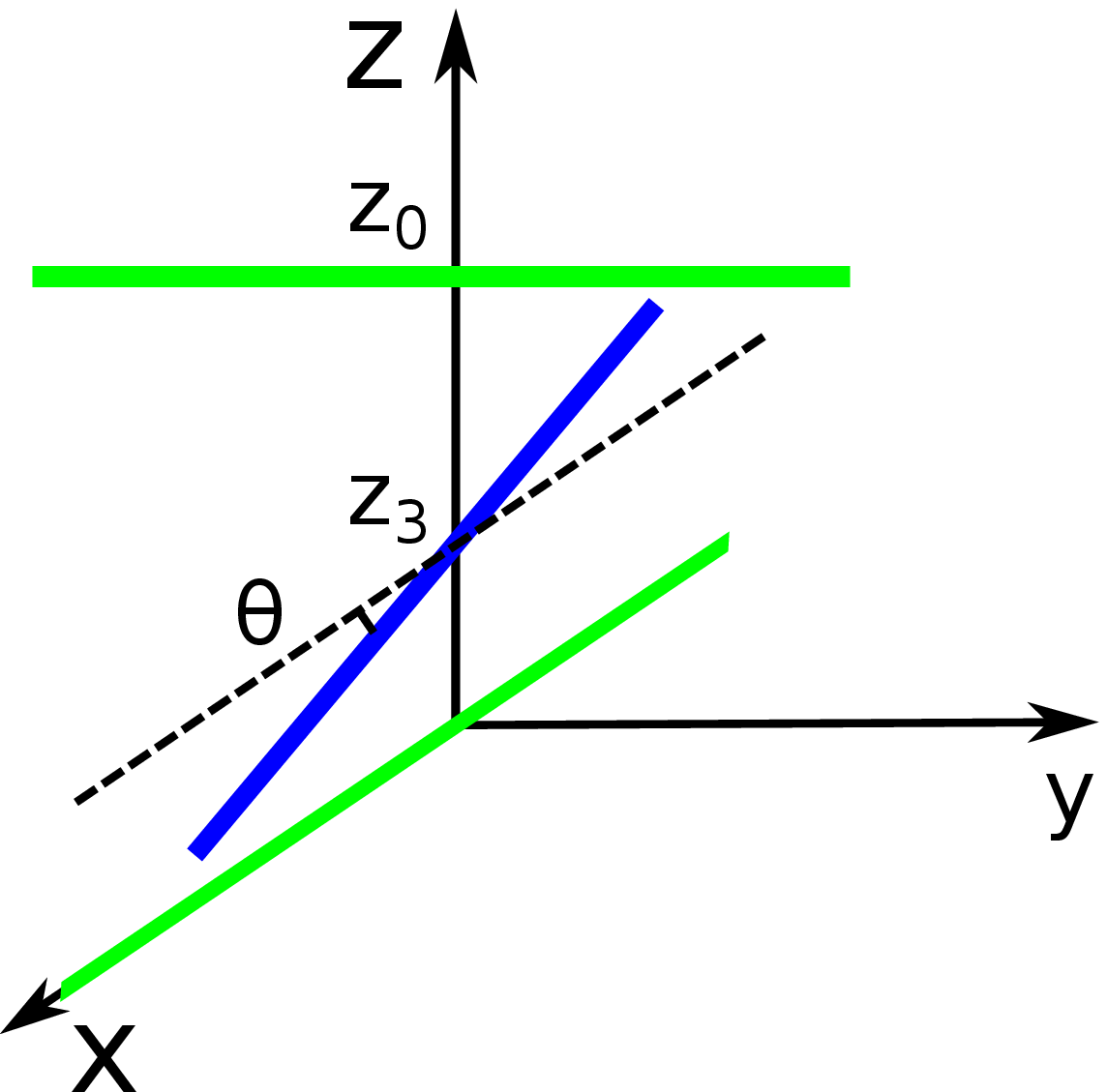}}
\caption{(Color online) (a) A schematic plot of crystallized bundles in a network, as excerpted
  from Ref.~\cite{cui2010spontaneous}. Reprinted with permission from AAAS. (b)
  A schematic plot of the geometric constraint surrounding a filament
  represented by the middle blue rod.  }
\label{schematic}
\end{figure}

We model the interaction between a bundle and its surrounding filaments by a
geometric constraint and a confinement potential. In experiment, the bundles are
randomly oriented and their interlocking in the network limits the mobility of
any bundle within some channel around them, as schematically shown in
Fig.~\ref{schematic}(a)~\cite{hartgerink2001self}. The shape of the channel is
assumed to be circular here. This geometric constraint is represented by a
hard-wall potential. Furthermore, filaments in a bundle are subject to a
confinement potential arising from the electrostatic repulsion between the
filaments and the wall. To obtain the expression for the confinement potential,
we first calculate the Coulomb interaction energy of a single filament in an
arbitrary bundle in a filament network. Filaments in neighboring bundles are
represented by the two green lines at $z=0$ and $z_0$ in Fig~\ref{schematic}(b). For filament
length $20\ \mu \textrm{m}$, $z_0=100\ \textrm{nm}$ and $\theta =\pi/4$,
numerical calculations show that the potential energy of a charged filament
between two perpendicular ones versus its position $z_3$ can be well fitted by a
quadratic curve. The collective interactions from all filaments in neighboring
bundles enhance the potential energy of the charged filament (the blue one at
$z=z_3$) without modifying the quadratic law; the sum of quadratic polynomials
is also a quadratic polynomial.  Based on the above heuristic
calculation, the confinement potential is assumed to conform to a quadratic law
in the non-screening regime. Considering the screening effect of solutions, we
model the influence of the wall on a filament as decaying exponentially. The
expression for the confinement potential must reduce to a quadratic form as the
screening length approaches infinity. We therefore propose the expression for
the confinement potential as \begin{eqnarray} V_{conf}(r) = \beta
  \left(\frac{r}{R}\right)^2 \exp\left(-\kappa (R-r) \right) ,\label{Vconf}
\end{eqnarray} where $r$ is the distance from the center of the channel to a
filament. Here we introduce the phenomenological parameter $\beta$ to
characterize the strength of the confinement potential. It has a complicated
dependence on the charge density of filaments as well as their orientations and
positions in the network. It is interesting to compare Eq.(\ref{Vconf}) with the
confinement potential between two quarks that can be approximated by $B r
e^{br}$ (both $B$ and $b$ are constants)~\cite{hsu1981confinement}. Note that
the optimal angle $\theta$ defined in Fig~\ref{schematic}(a) is calculated to be
always $\pi/4$ with the position $z_3$ of the blue line varying between $0.1
z_0$ and $0.5 z_0$. This result supports the hypothesized templating effect in
the formation of networks, which states that long filaments formed at early
stages act as templates for the formation of bundles as the growth of short
filaments continues~\cite{hartgerink2001self}.

To summarize the above discussion, the energetics of $N$ particles in a
disk of radius $R$ representing $N$ filaments in a bundle is: \begin{eqnarray}
  f[\{ \vec{r}_i \}] = \alpha \sum_{i=1}^N H(||\vec{r}_i||-R)+ \sum_{i=1}^N
  V_{conf}(||\vec{r}_i||) \\ + \sum_{i\neq j}
  V_{int}(\vec{r}_i-\vec{r}_j) \nonumber ,\label{Eexpression}\end{eqnarray} where $\vec{r}$
  is the two-dimensional position vector of a particle in a disk. The
  first two terms are the hard-wall potential and the confinement potential,
  respectively. $H(x)$ is the Heaviside step function; it is zero for $x<0$ and
  1 for $x\geq 0$. The parameter $\alpha$ is a large number characterizing the
  hard-wall
  potential. The last term in Eq.(3) describes the interaction between
  filaments. Note that in the limit of large screening length,
  Eq.(3) is recognized as the constrained two-dimensional Coulomb
  gas model~\cite{mastatistical}. For an electrically neutral network, only the geometric
  constraint term in Eq.(3) survives. The confinement potential
  term tends to push particles towards the center of the disk, while the
  particle-particle repulsion term prevents their approach.

  We perform annealing Monte Carlo
  simulation for identifying the lowest-energy configuration of particles
  confined in a disk~\cite{newman1999monte}. About $10^6$ MC sweeps are carried out for each run; each MC
  sweep consists of trial attempts to randomly move each particle. 
The acceptance or rejection of a MC trial
is determined by the standard Metropolis algorithm. The hard-wall potential
is treated as a geometric constraint, i.e., the
particles are not allowed to move beyond the disk boundary. In the
simulation, the functional to be minimized is $\tilde{f}[\{ \vec{r}_i \}] =
\Gamma \sum_i (\frac{||\vec{r}_i||}{R})^2 \exp\left(-\kappa (R-||\vec{r}_i||)\right)
 + \sum_{i\neq j} K_0 \left( \kappa\ ||\vec{r}_i-\vec{r}_j|| \right)$, which
 reduces to $\tilde{f}_{Coulomb}[\{ \vec{r}_i \}] =
\Gamma \sum_i (\frac{||\vec{r}_i||}{R})^2-\sum_{i\neq j}
\ln(||\vec{r}_i-\vec{r}_j||)+\textrm{const}$ in
the limit of large screening length.
The phenomenological dimensionless parameter $\Gamma$ controls the relative importance of the
confinement potential and the interaction between particles. In simulation, we
set $R=1$ which defines a unit length. Other length scales are measured in
terms of the radius of the disk.

In oder to characterize the hexagonal crystalline order, we construct bonds between
particles via the Delaunay triangulation~\cite{de2008computational} and
introduce a bond order parameter $|\Phi_6|^2$ on the constructed triangular
lattice ~\cite{nelson1979dislocation} \begin{eqnarray} \Phi_6 =
  \frac{1}{N}\sum_{m=1}^N \frac{1}{N_b} \sum_{n=1}^{N_b}\exp\left(6 i
  \theta_{mn} \right),\label{Phi6}\end{eqnarray} where $\theta_{mn}$ describes
  the orientation of the bond connecting the two neighboring particles $m$ and
  $n$ relative to some fixed reference axis.  The modulus of $\Phi_6$ is
  independent of a global rotation of the system. $|\Phi_6|^2=1$ for a perfect
  hexagonal crystal and $|\Phi_6|^2=0$ for a liquid state. For eliminating the
  edge effect in a finite system, the exterior particles are excluded in the
  calculation of the order parameter $|\Phi_6|^2$.

\begin{figure}[t]  
 \centering 
\includegraphics[width=2in]{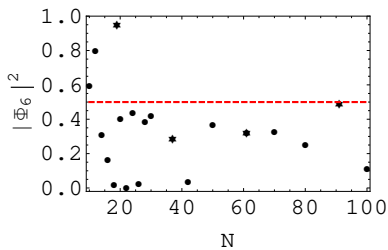}
\caption{(Color online) The order parameter $|\Phi_6|^2$ versus the number of filaments $N$. The
  stars are corresponding to centered hexagonal numbers (19, 37, 61 and 91). Measured in
terms of the disk radius $R$, $\kappa^{-1} = 0.1$. $\Gamma = 5$. $R=1$. }
\label{psi_N}
\end{figure}

\begin{figure}[h]  
\centering 
\subfigure[]{
\includegraphics[width=0.8in]{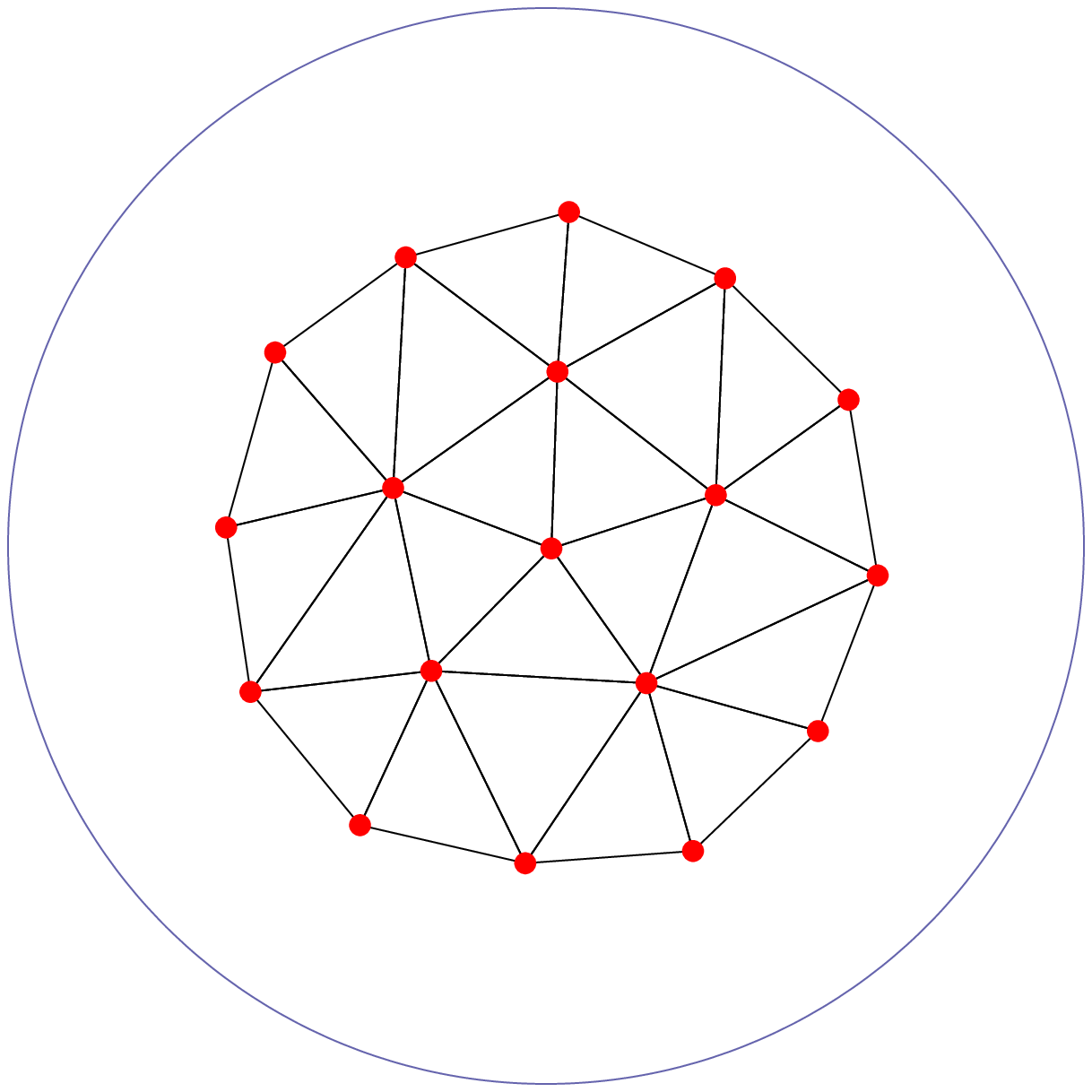}}
\hspace{-0.05in} 
\subfigure[]{
\includegraphics[width=0.8in]{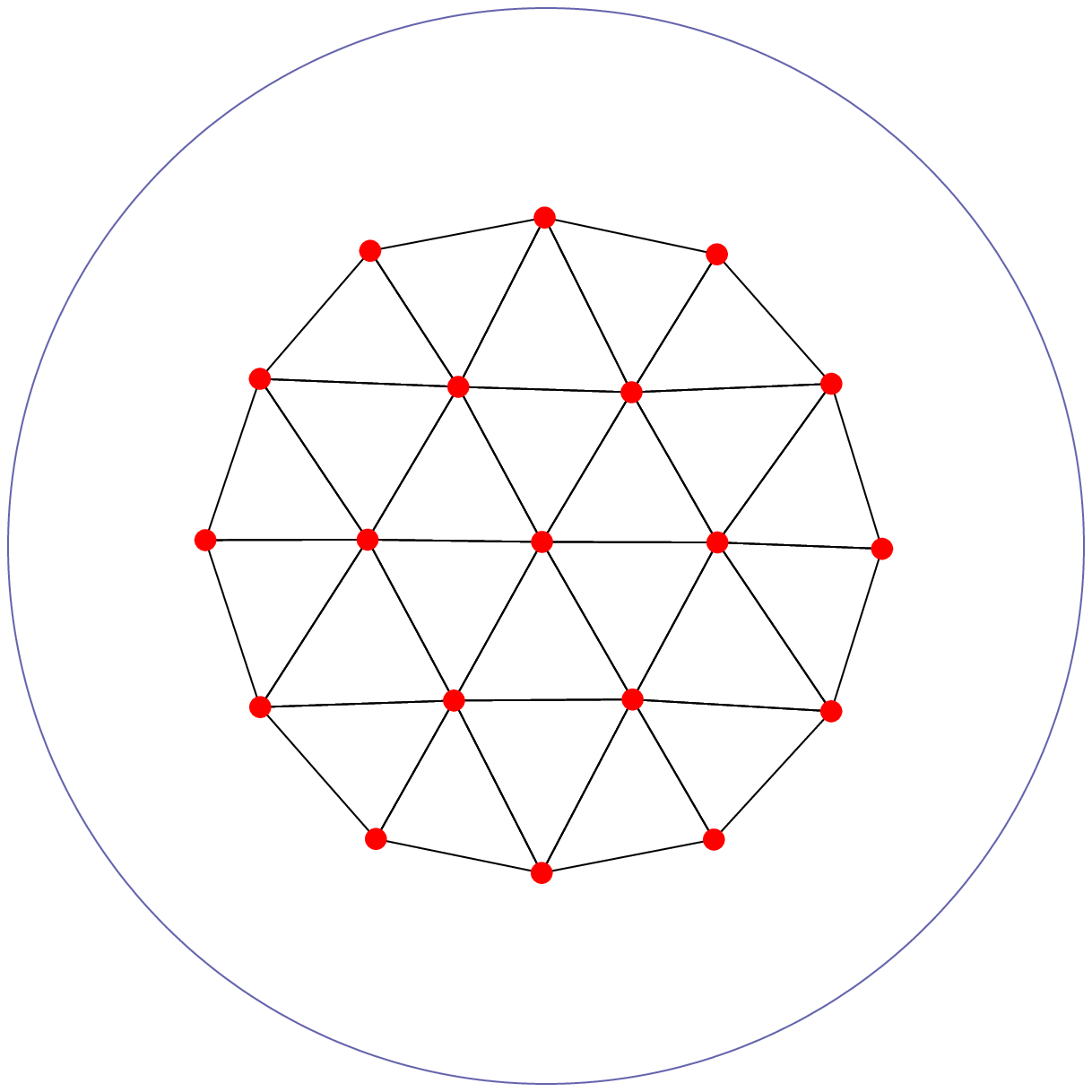}}  
\hspace{-0.05in} 
\subfigure[]{
\includegraphics[width=0.8in]{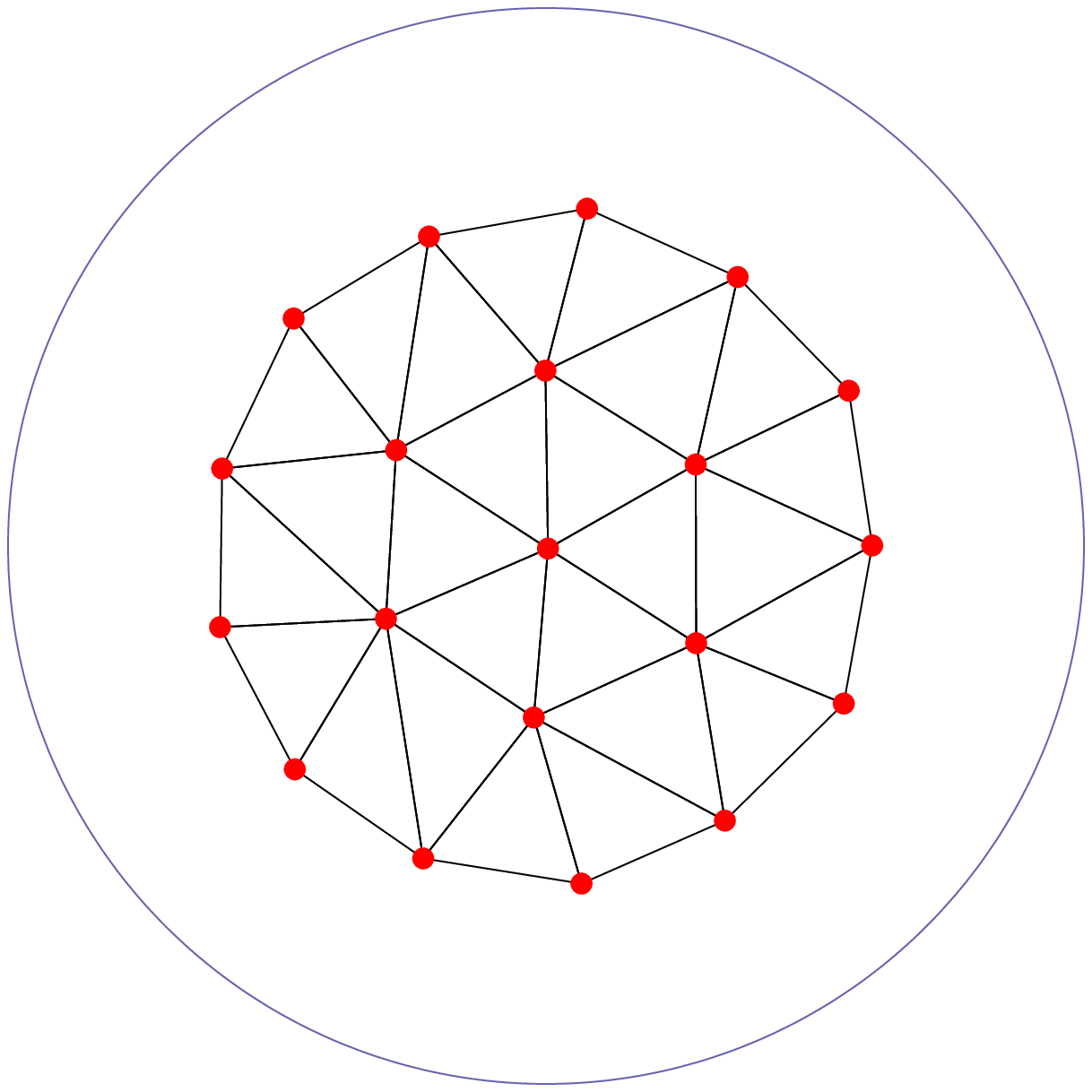}}
\hspace{-0.05in} 
\subfigure[]{
\includegraphics[width=0.8in]{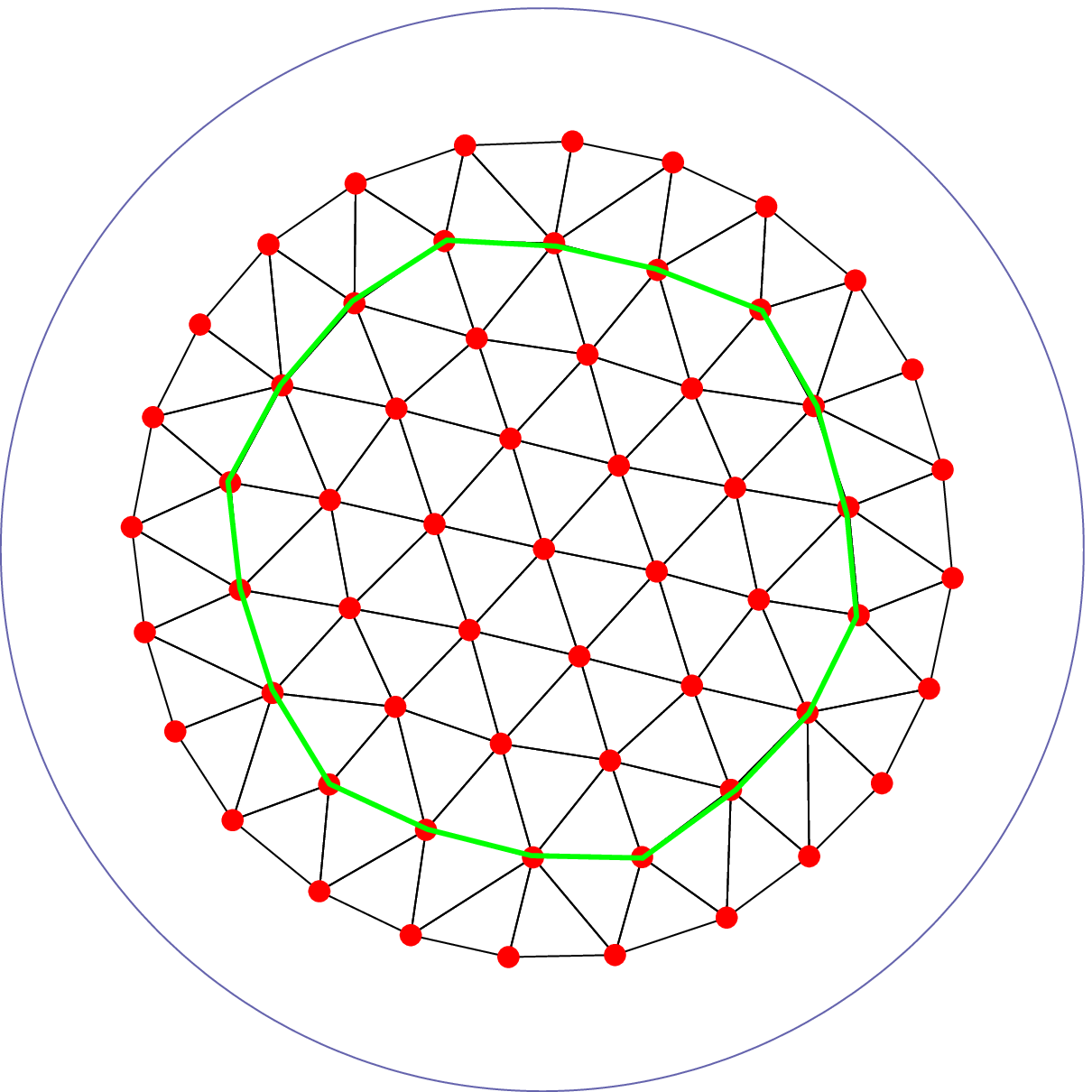}}
\hspace{-0.05in} 
\subfigure[]{
\includegraphics[width=0.8in]{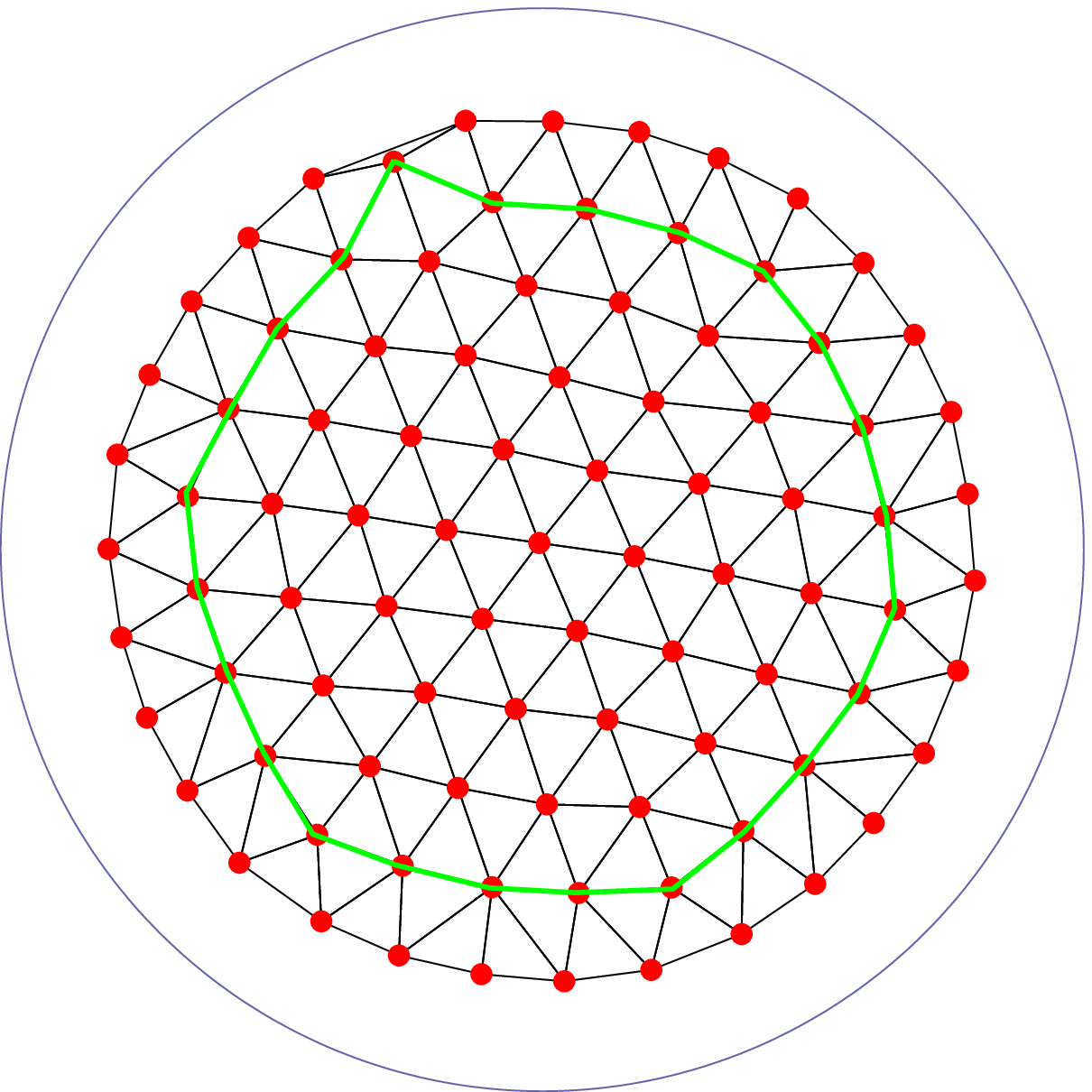}}
\hspace{-0.05in} 
\subfigure[]{
\includegraphics[width=0.8in]{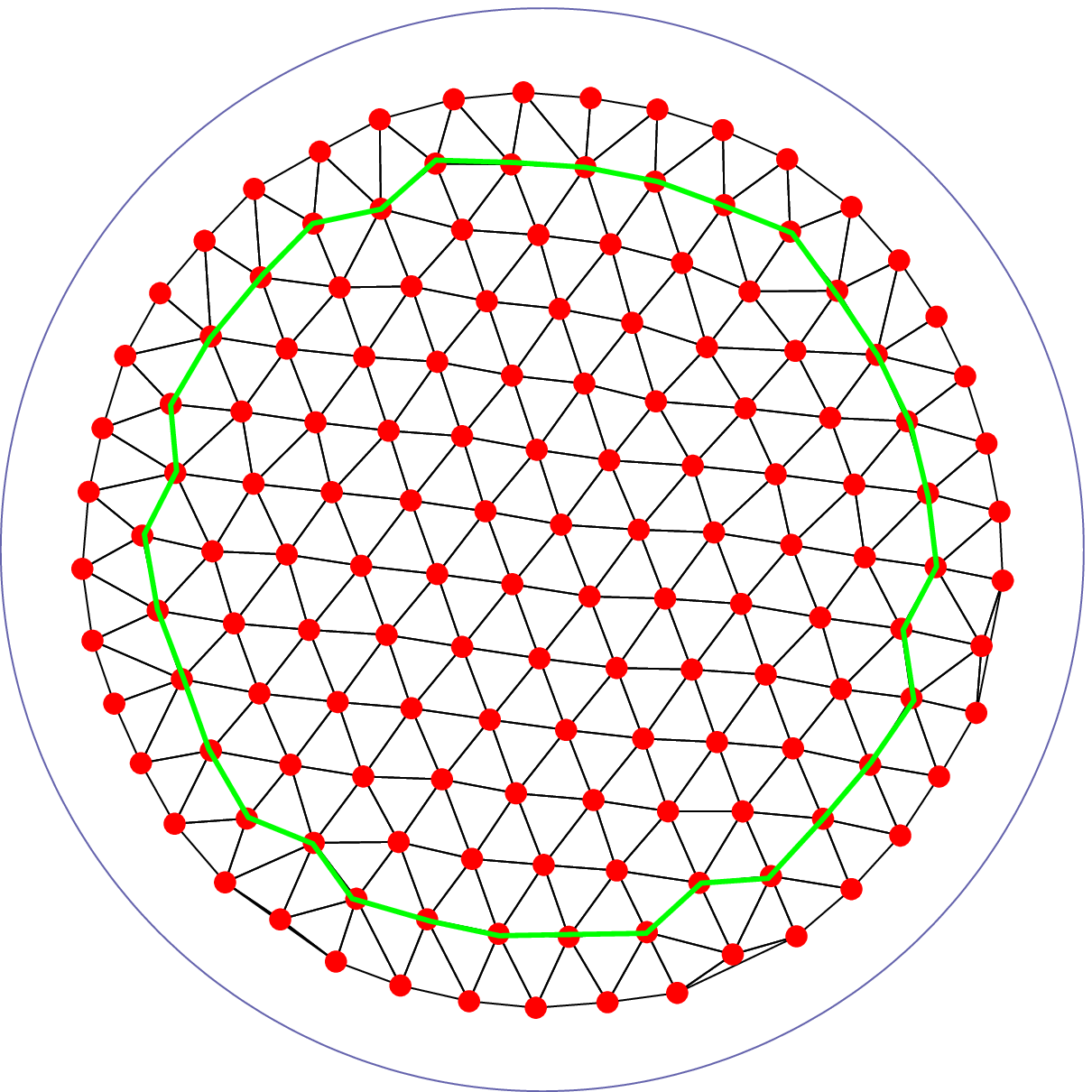}}
\caption{(Color online) The low-energy configurations of $N$ filaments represented by red dots
  that are confined in a bundle subject to the confinement potential
  $V_{conf,sc}$. The particles are connected via the Delaunay triangulation. The
  blue circle represents the hard-wall. $N = 18$ (a), $19$
  (b), $20$ (c), $61$ (d), $91$ (e) and $150$ (f). Measured in
terms of the disk radius $R$, $\kappa^{-1} = 0.1$. $\Gamma = 5$. $R=1$. }
\label{config_N}
\end{figure}

\section{Results and Discussion}

In experiments, the hexagonal crystalline order emerges in bundles of filaments
with the increase of charges on filaments~\cite{cui2010spontaneous, stark1991effect}. In this
process, the parameter $\Gamma$, which characterizes the relative strength of the
confinement potential to the interaction between particles, varies
correspondingly. The bundle size is rather polydispersed; the number of filaments $N$ in a bundle
is in the magnitude of $10-100$. We systematically study bundles of varying
sizes. Figure~\ref{psi_N} shows that the dependence of $|\Psi_6|^2$ on $N$ is
highly non-monotonous. For example, $|\Psi_6|^2 = 0.95$ for $N=19$, and it
suddenly drops
to $0.02$ or $0.4$ by decreasing or increasing one particle to the system. This phenomenon can
be attributed to the geometric specialty of the number 19. It is a centered
hexagonal number. A centered hexagonal number $N_{hex}$ is the number of a hexagon with a dot at the
center and all other dots surrounding the central dot in a hexagonal lattice.
Adding or removing a point from a perfect hexagonal lattice composed of $N_{hex}$
would destroy the perfect crystalline structure. This phenomenon is shown in
Fig.\ref{config_N} (a-c); $N =$ 18, 19 and 20 from (a) to (c). In Fig.~\ref{psi_N}, the points above the red line may be regarded as in a
crystallized state; the configurations of $N= 19$ is shown in
Fig.~\ref{config_N} (b). Those below the red line may be in partially
crystallized states. For example, the interior particles in the configurations
of $N=61,\ 91$ and $150$ are perfectly crystallized, as shown in
Fig.~\ref{config_N}(d-f). Their low values of $|\Psi_6|^2$ are due to the topological defects near the
boundary.

\begin{figure}[h]  
\centering 
\subfigure[]{
\includegraphics[width=0.8in]{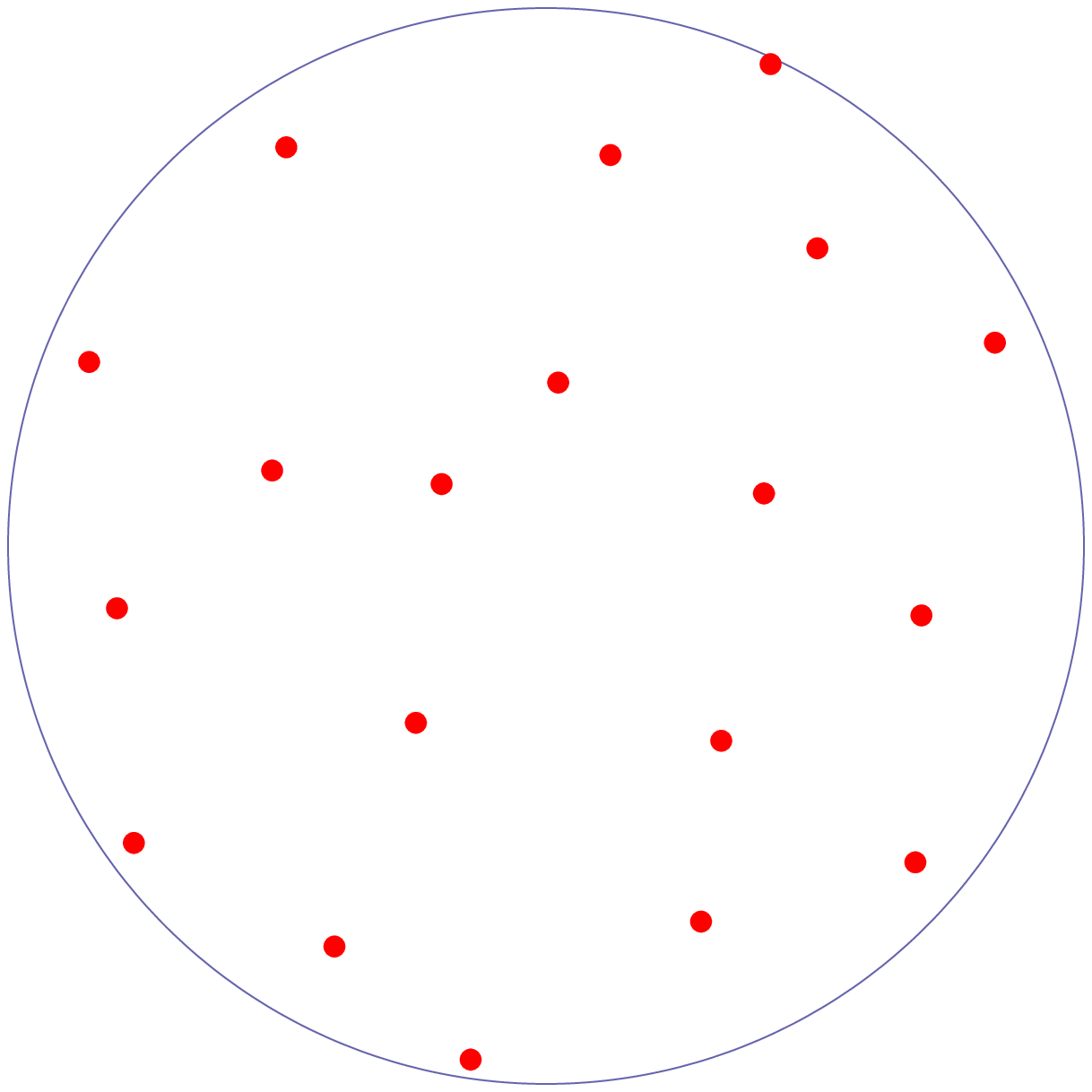}}
\hspace{-0.05in} 
\subfigure[]{
\includegraphics[width=0.8in]{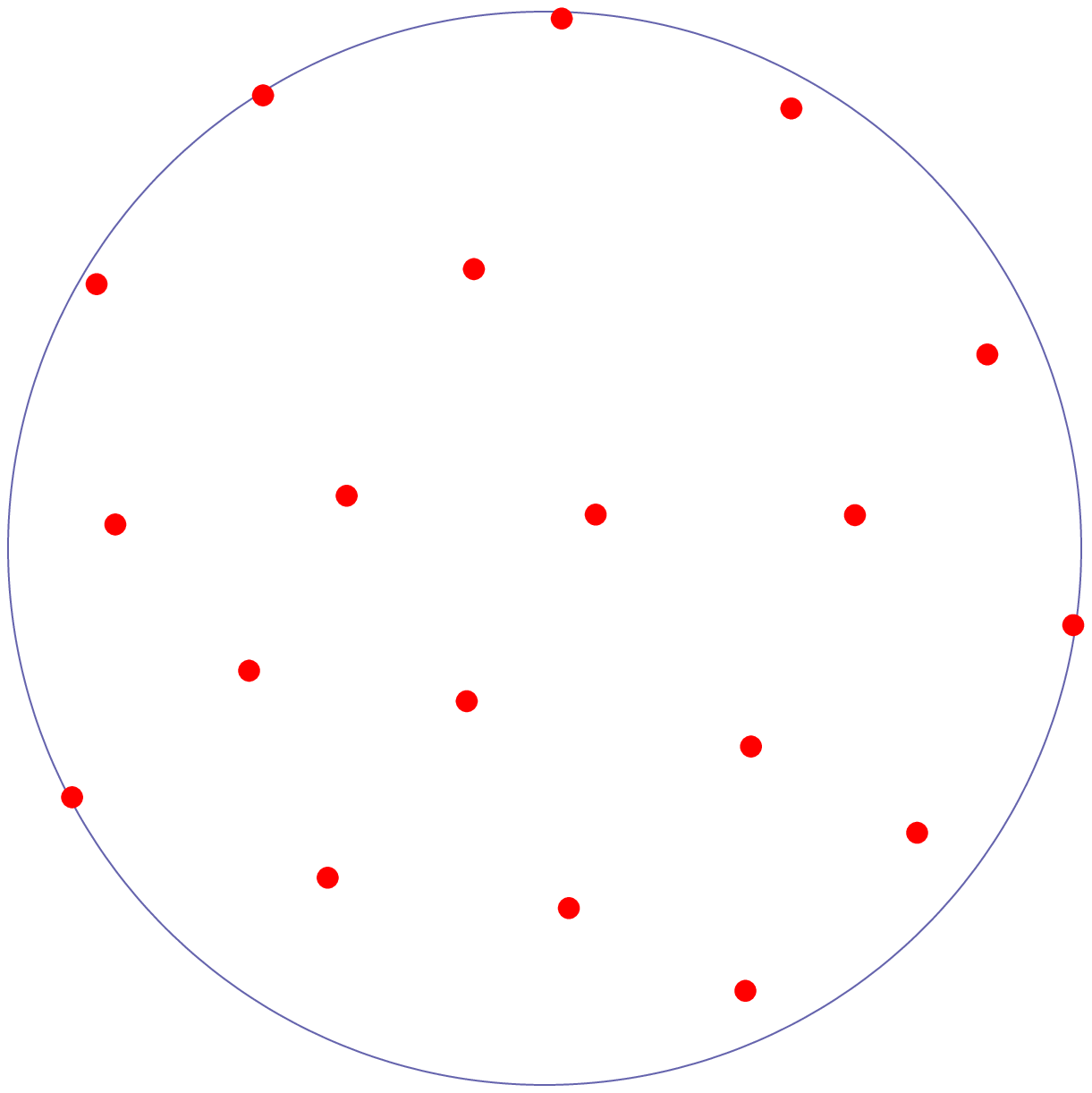}}  
\hspace{-0.05in} 
\subfigure[]{
\includegraphics[width=0.8in]{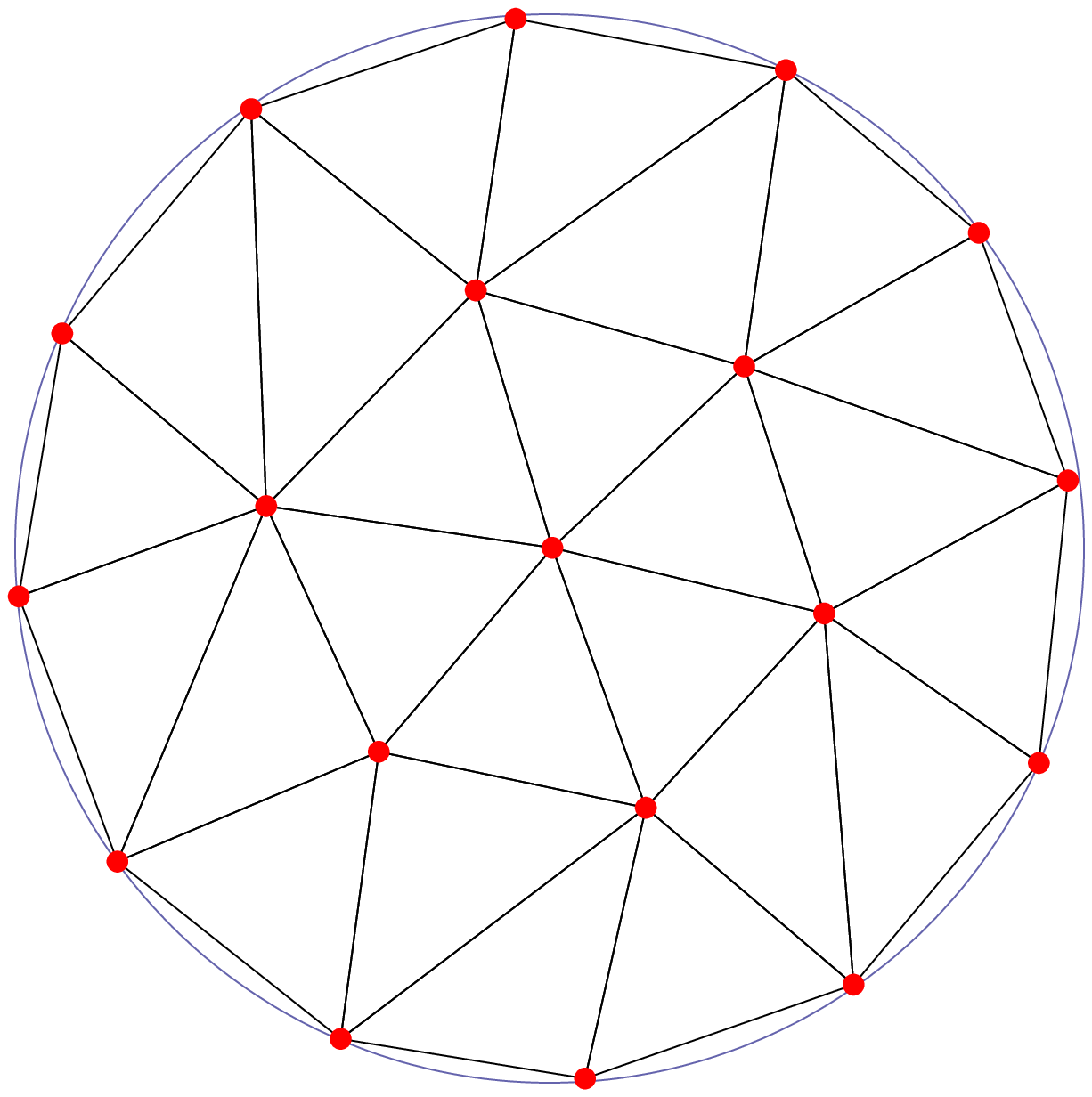}}
\hspace{-0.05in} 
\subfigure[]{
\includegraphics[width=0.8in]{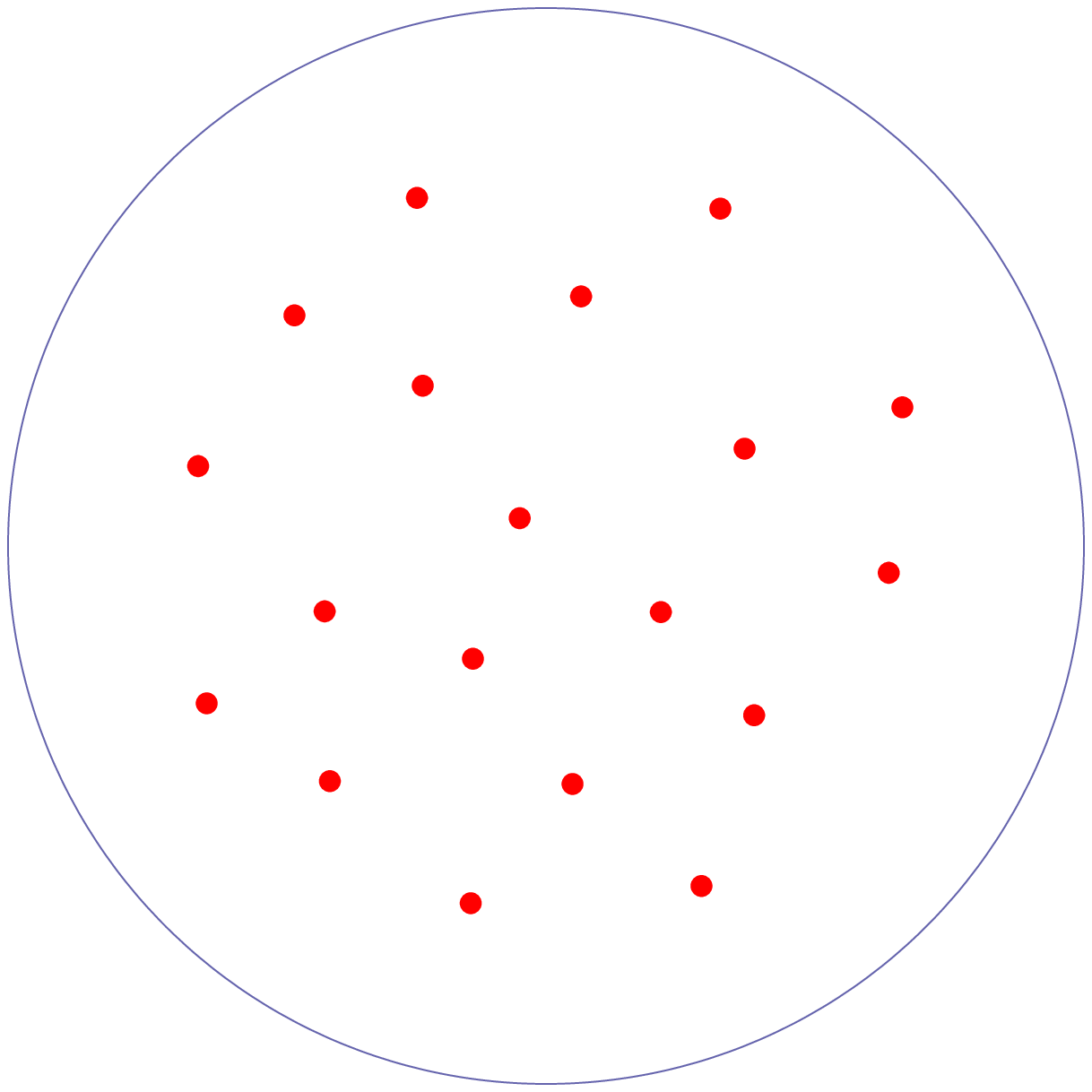}}
\hspace{-0.05in} 
\subfigure[]{
\includegraphics[width=0.8in]{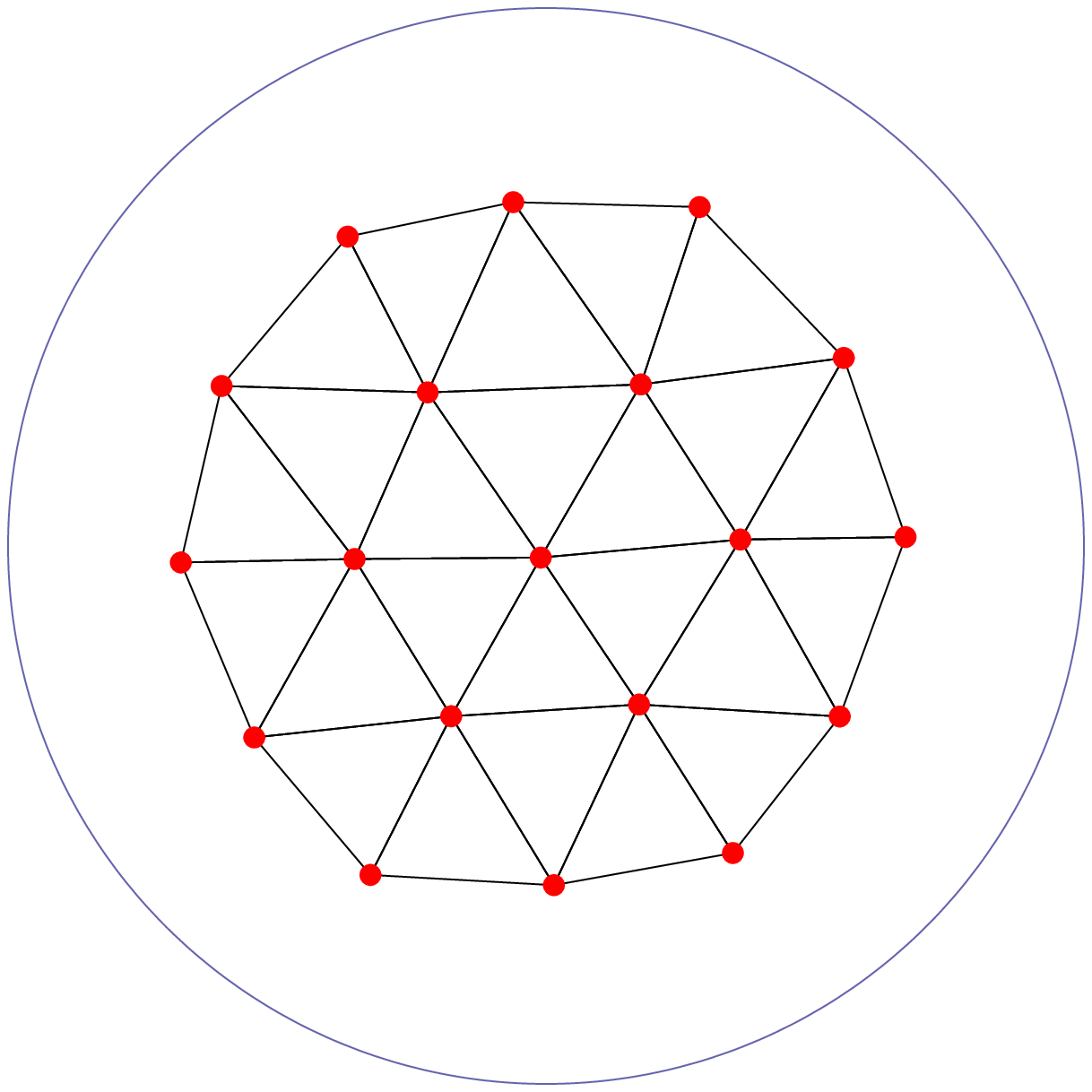}}
\hspace{-0.05in} 
\subfigure[]{
\includegraphics[width=0.8in]{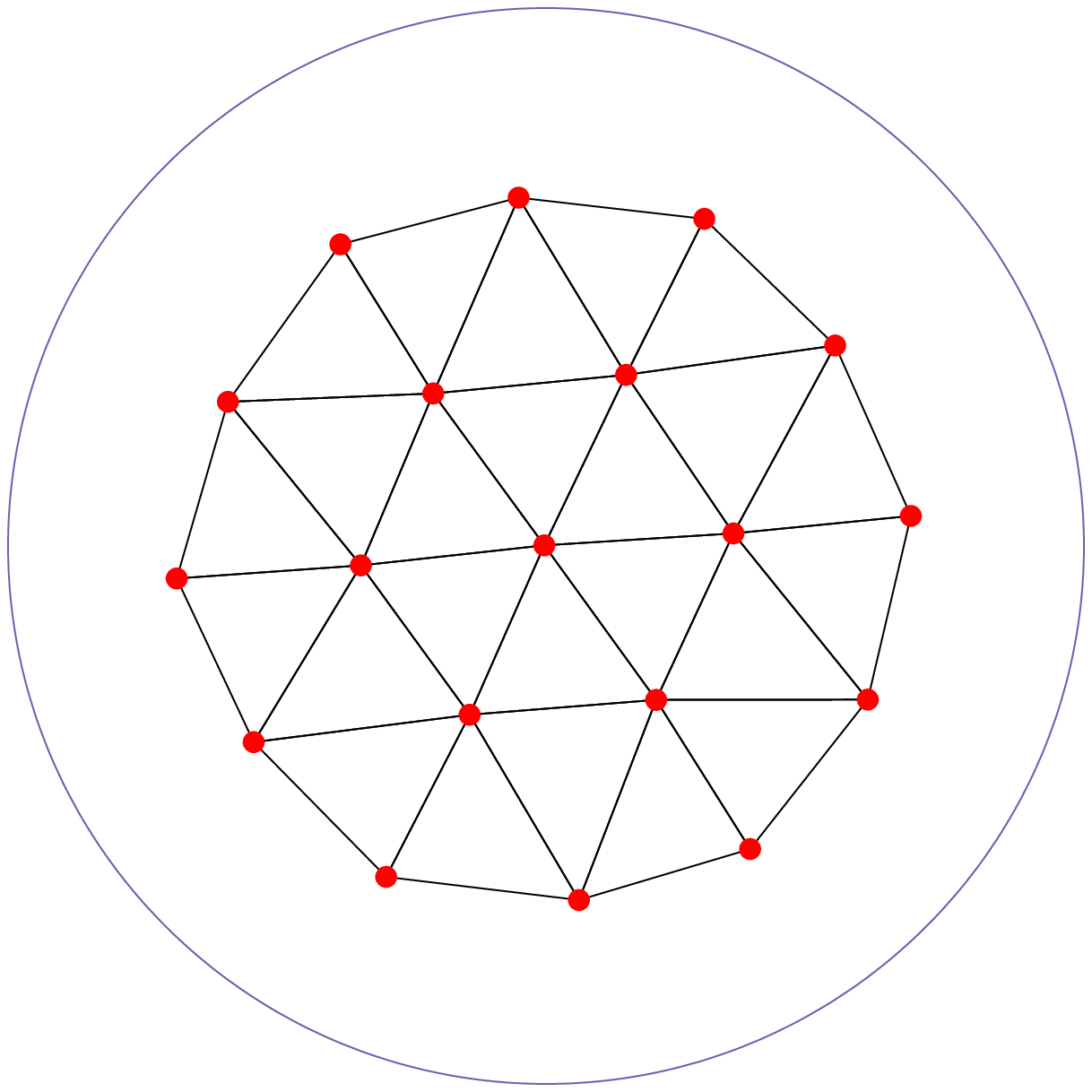}}
\caption{(Color online) The low-energy configurations of 19 filaments represented by red dots
  that are confined in a bundle subject to the confinement potential
  $V_{conf,sc}$. The blue circle represents the hard-wall. With the increase of
  the screening length $\kappa^{-1}$, the crystalline order emerges. The Delaunay
  triangulations are constructed on the hexagonal lattices. Measured in
terms of the disk radius $R$, $\kappa^{-1}= 0.03$ (a, d), $0.04$ (b, e) and
$0.05$ (c, f).
$\Gamma =0$ for the first row and $\Gamma=1.0$ for the second row. $R=1$. }
\label{config_Omega}
\end{figure}

\begin{figure}[h]  
  \centering 
\subfigure[]{
\includegraphics[width=1.66in]{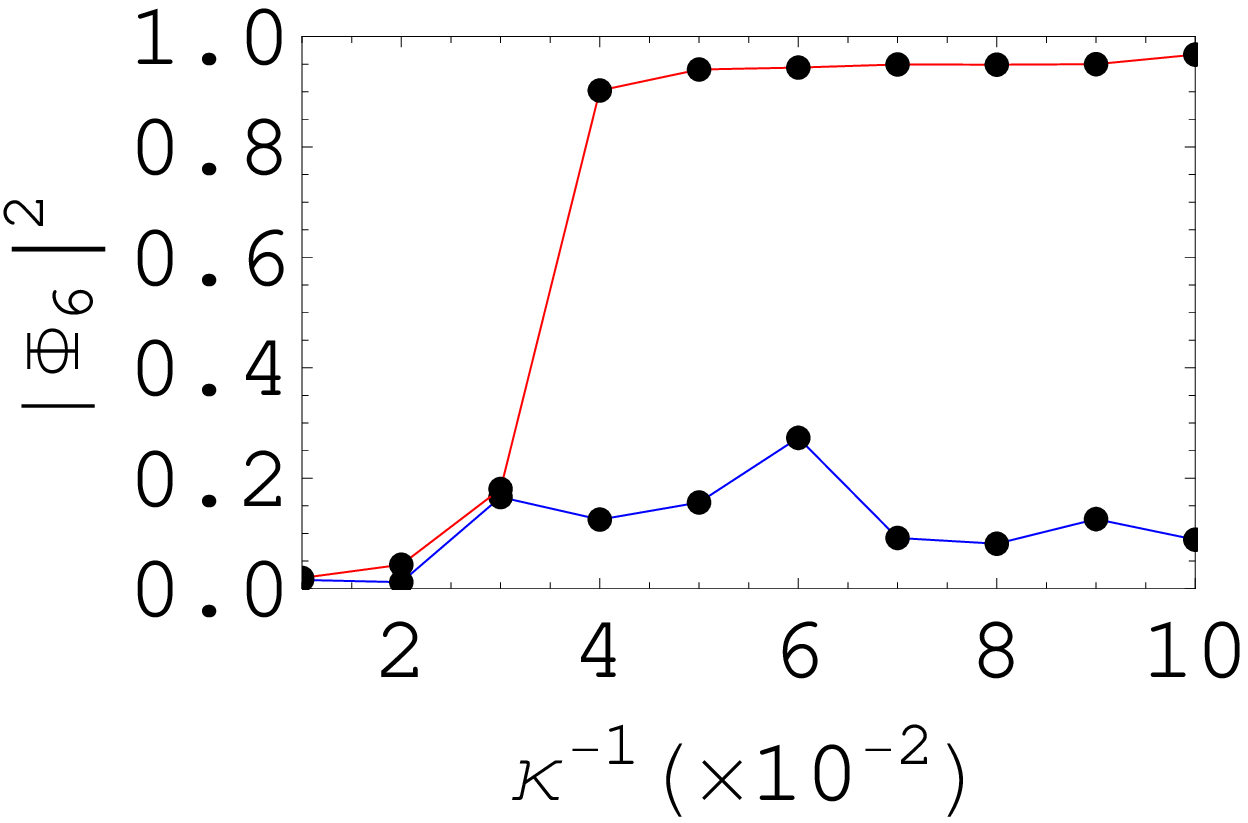}}
\hspace{-0.1in} 
\subfigure[]{
\includegraphics[width=1.66in]{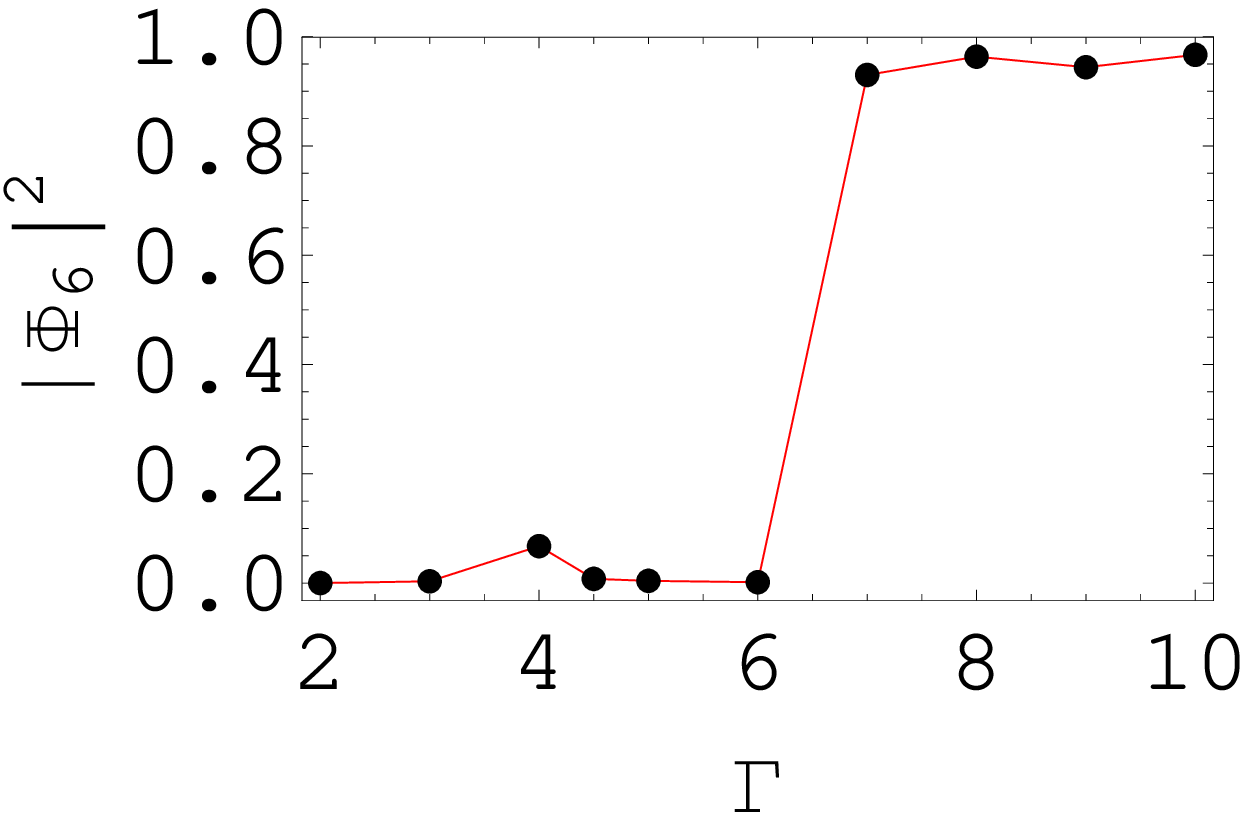}}
\caption{(Color online) (a) The order parameter $|\Phi_6|^2$ versus the screening length
  $\kappa^{-1}$. $\Gamma=0$ (the lower blue curver) and $1$ (the upper red curve). (b) $|\Phi_6|^2$ versus the
  parameter $\Gamma$ in the large
  screening length limit. The dots are corresponding to the configurations in Fig.~\ref{config_Gamma}. $N=19$. $R=1$. }
\label{phi_Gamma}
\end{figure}

\begin{figure}[h]  
\centering 
\subfigure[]{
\includegraphics[width=0.66in]{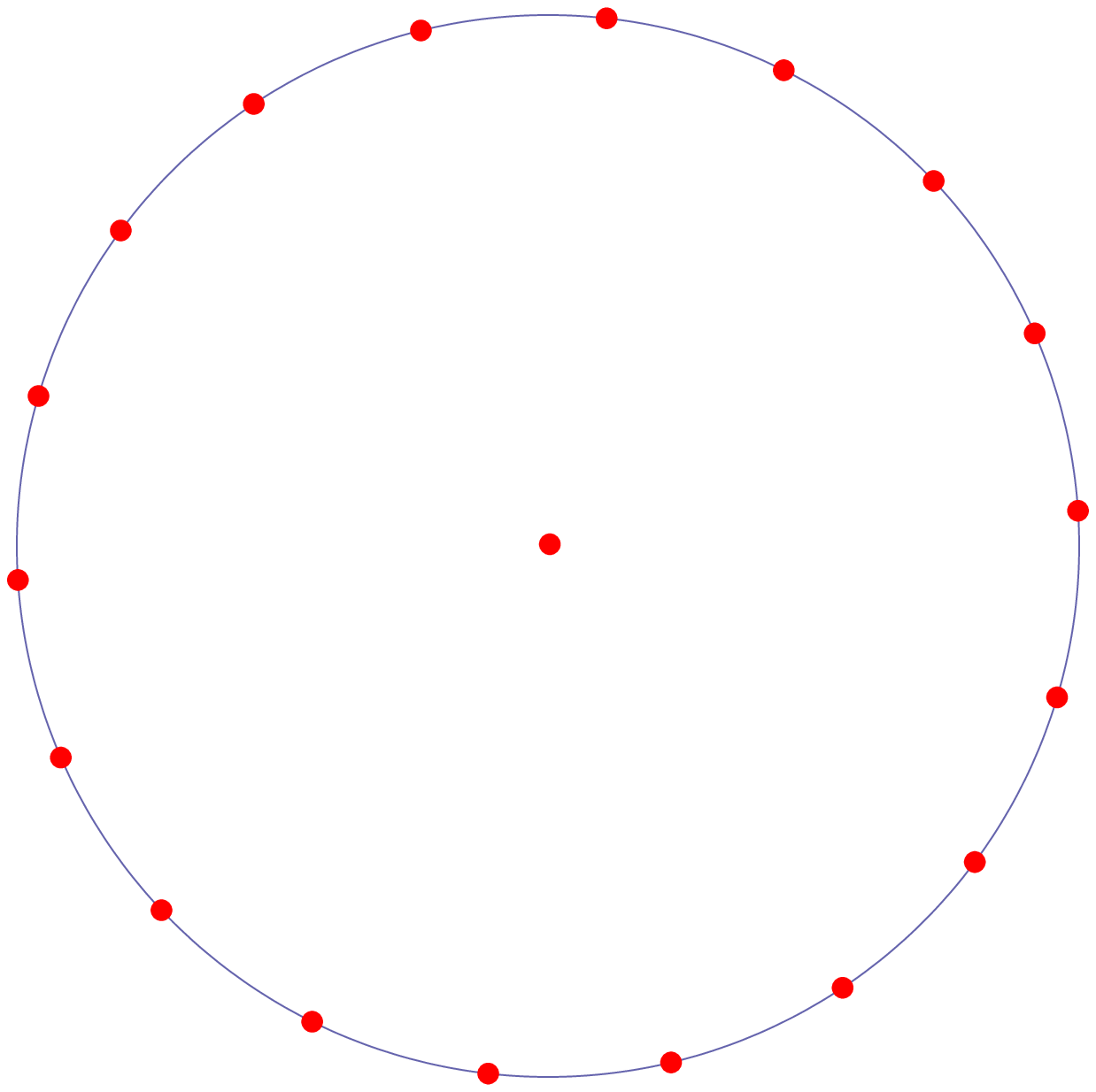}}
\hspace{-0.12in} 
\subfigure[]{
\includegraphics[width=0.66in]{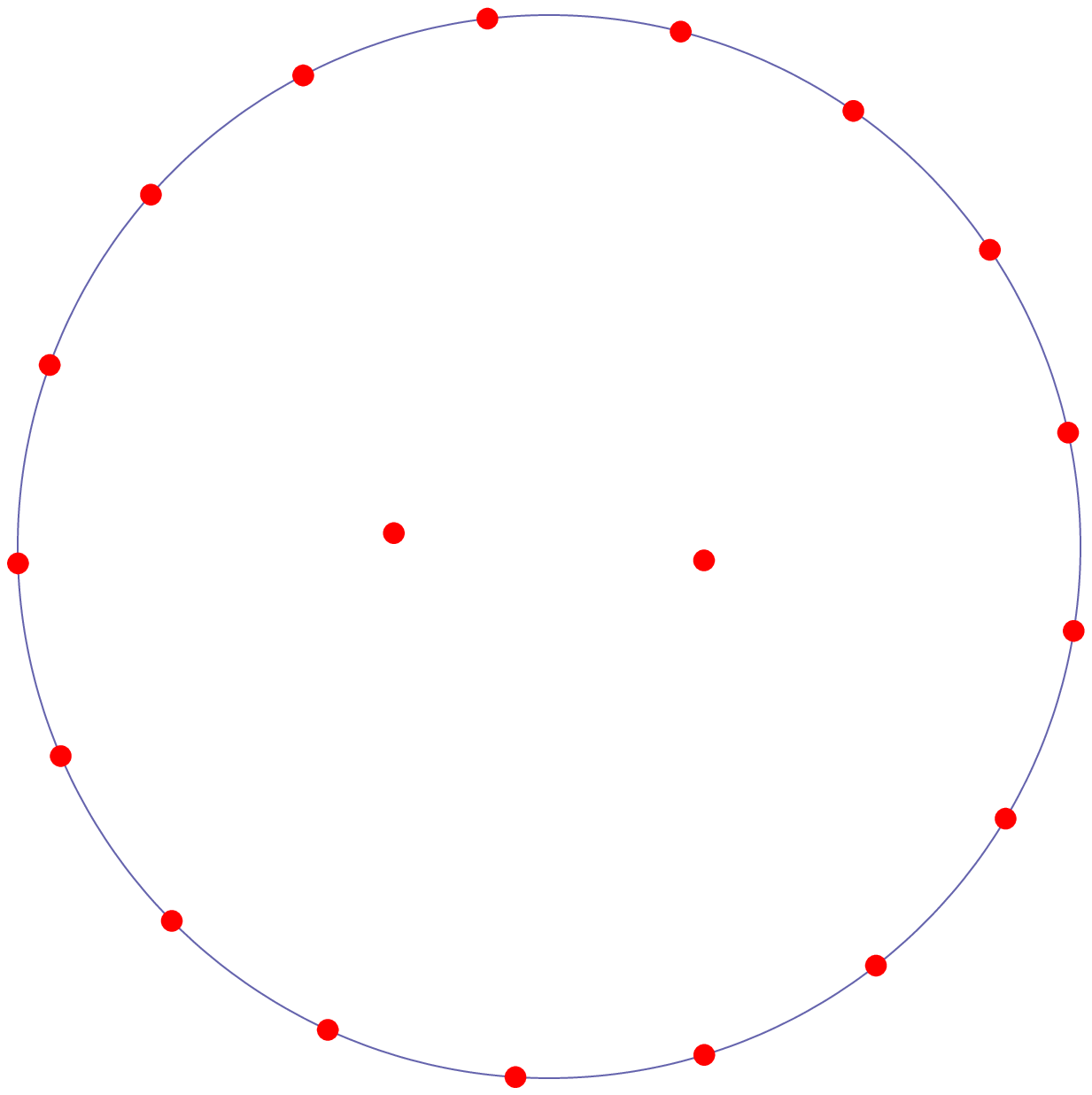}}  
\hspace{-0.12in} 
\subfigure[]{
\includegraphics[width=0.66in]{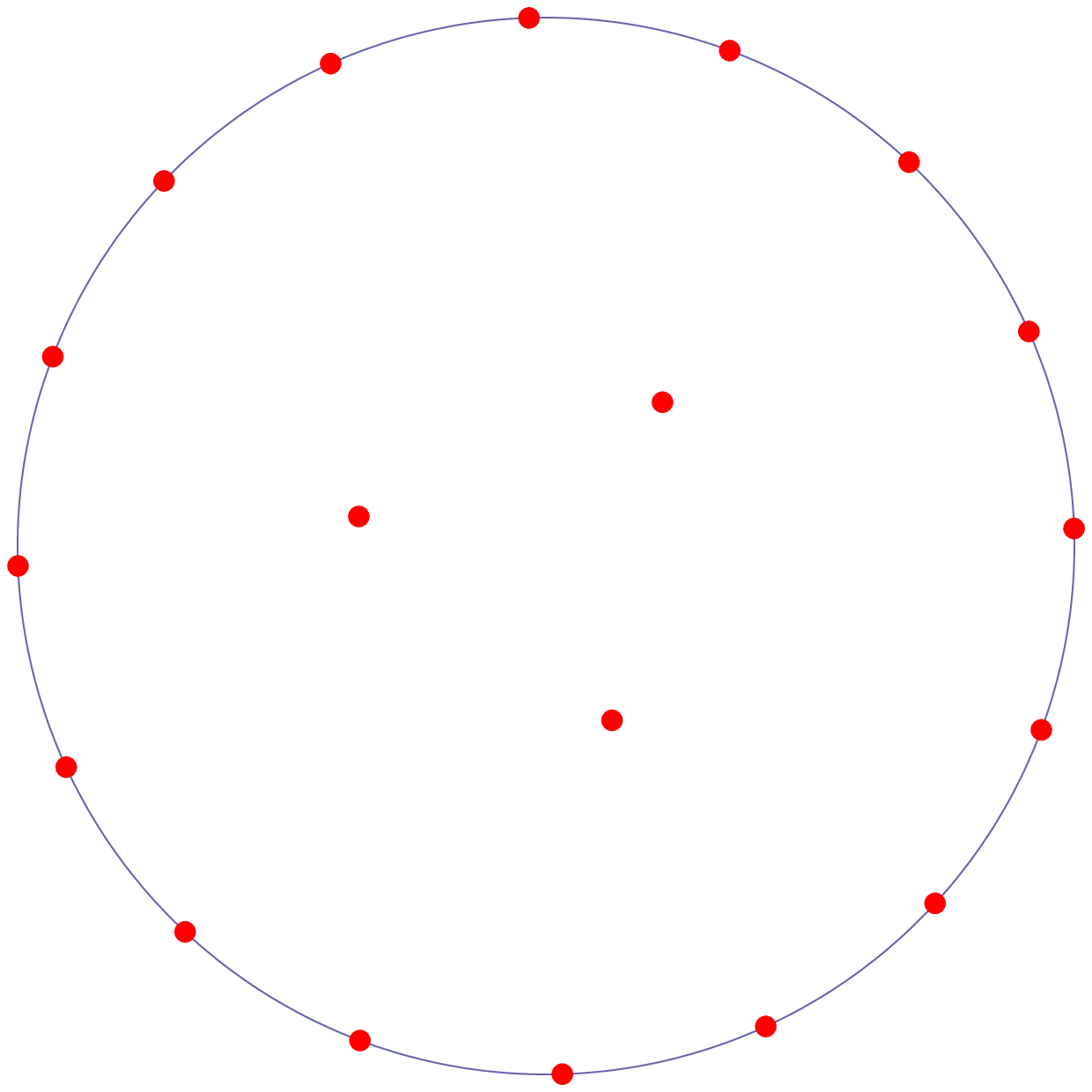}}
\hspace{-0.12in} 
\subfigure[]{
\includegraphics[width=0.66in]{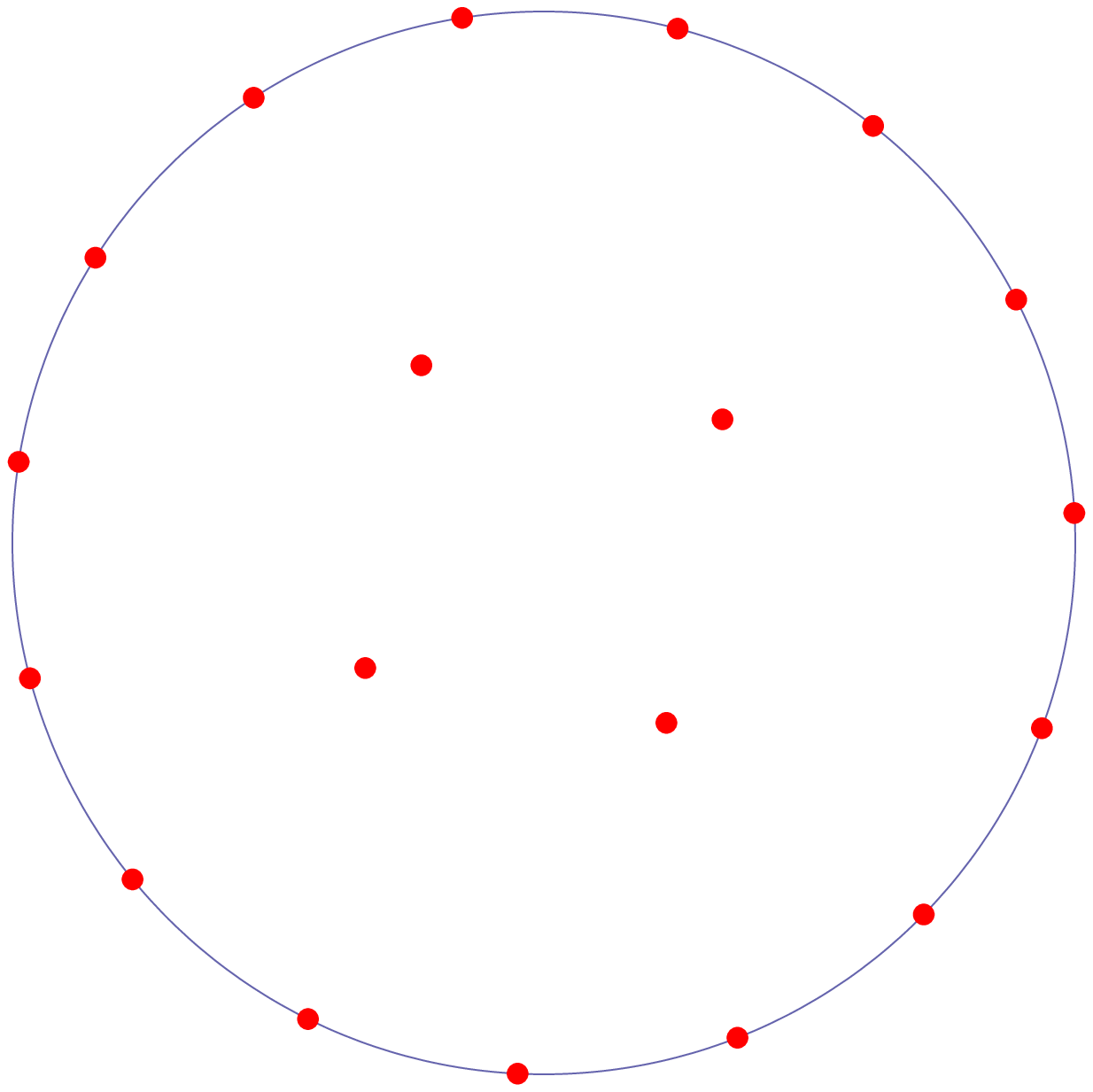}}
\hspace{-0.12in} 
\subfigure[]{
\includegraphics[width=0.66in]{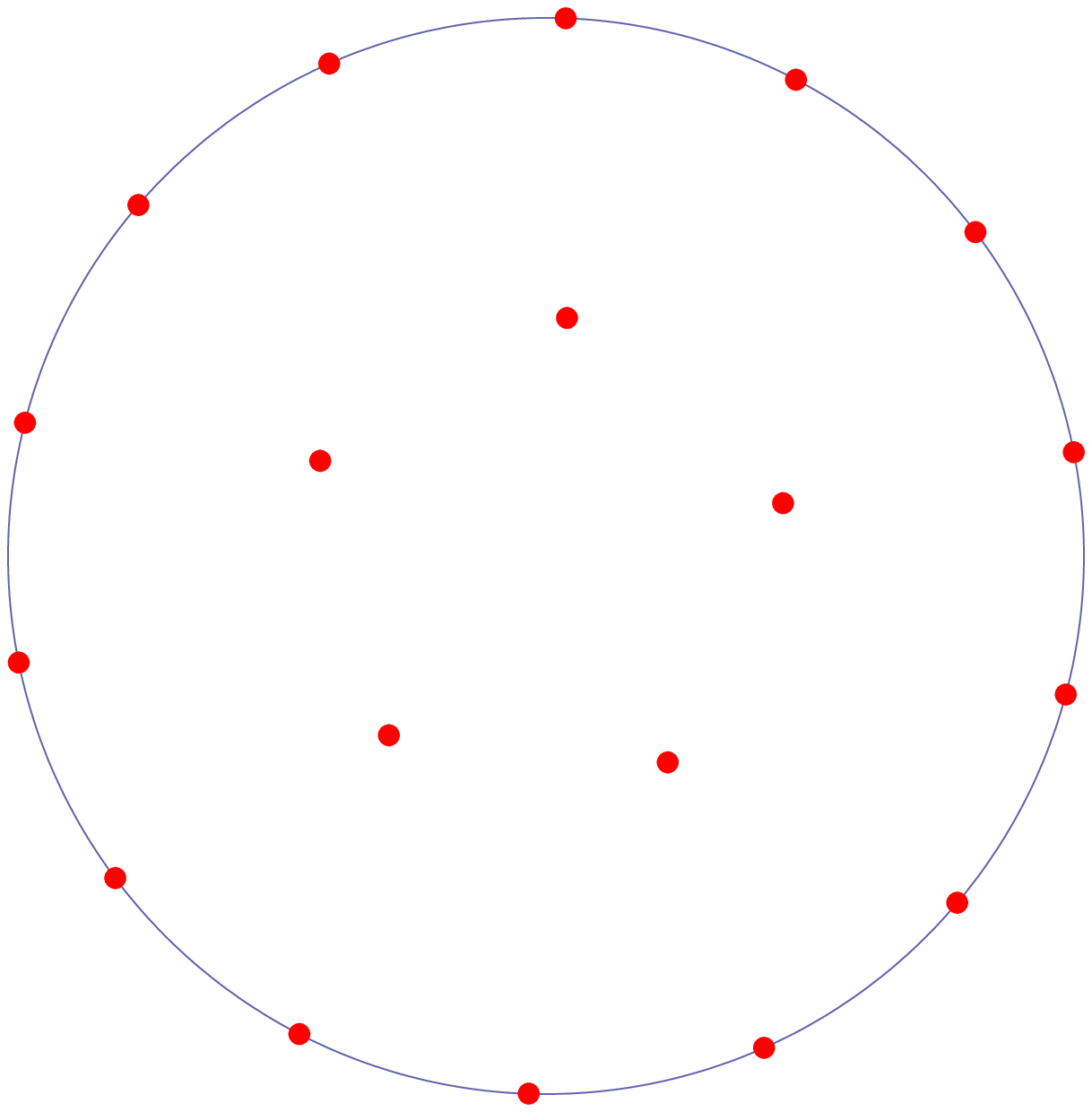}}
\hspace{-0.12in} 
\subfigure[]{
\includegraphics[width=0.66in]{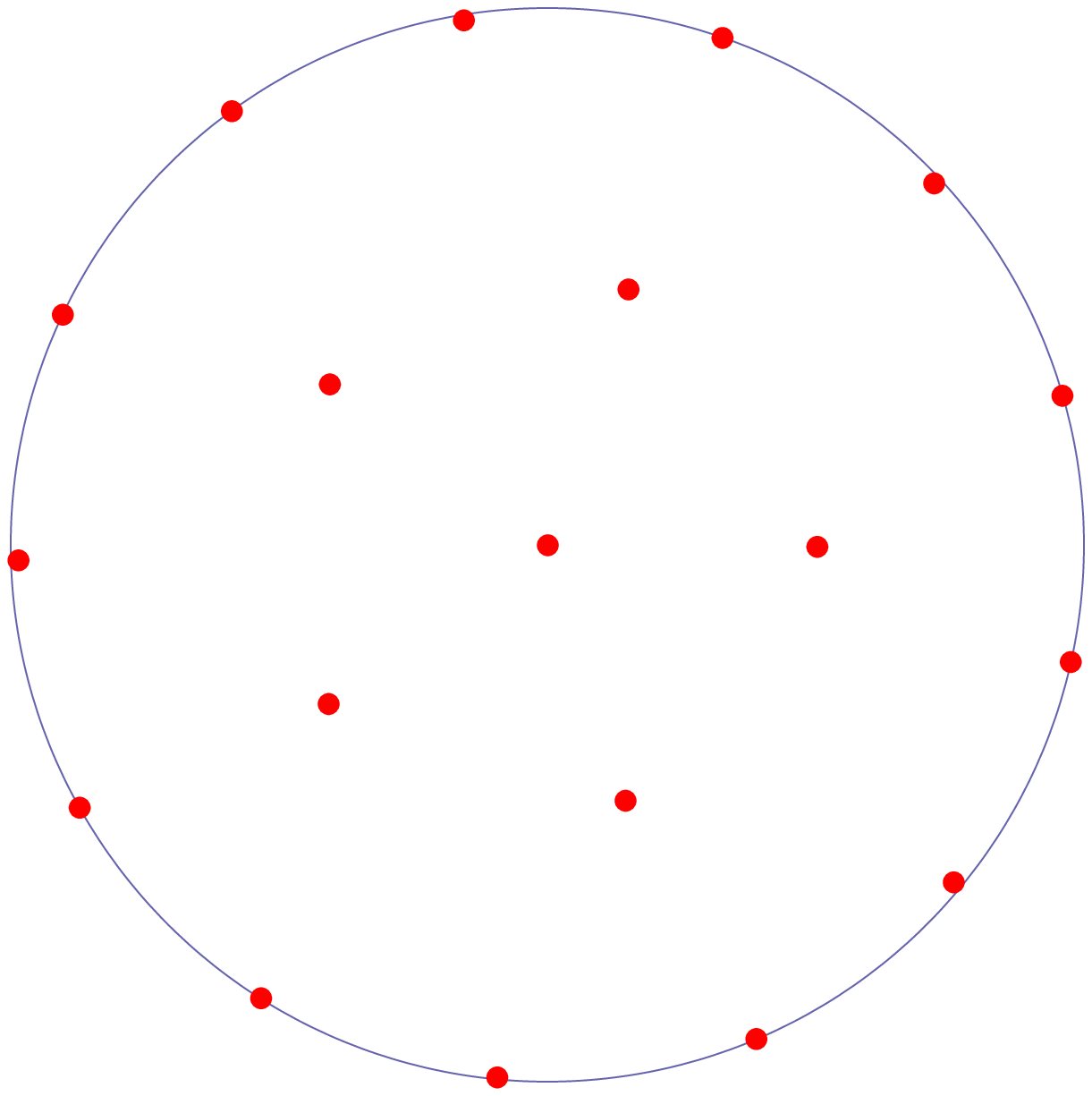}}
\hspace{-0.12in} 
\subfigure[]{
\includegraphics[width=0.66in]{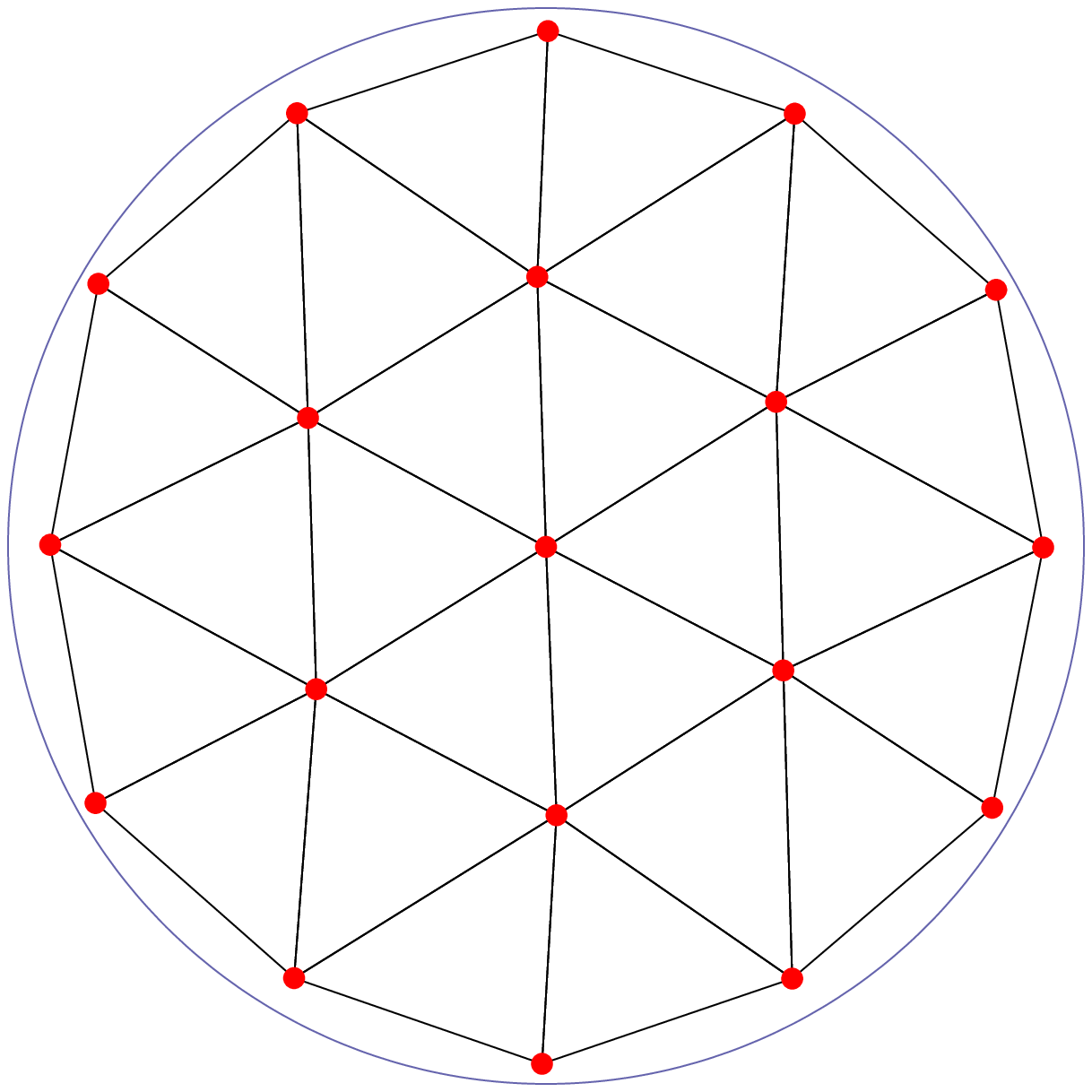}}
\hspace{-0.12in} 
\subfigure[]{
\includegraphics[width=0.66in]{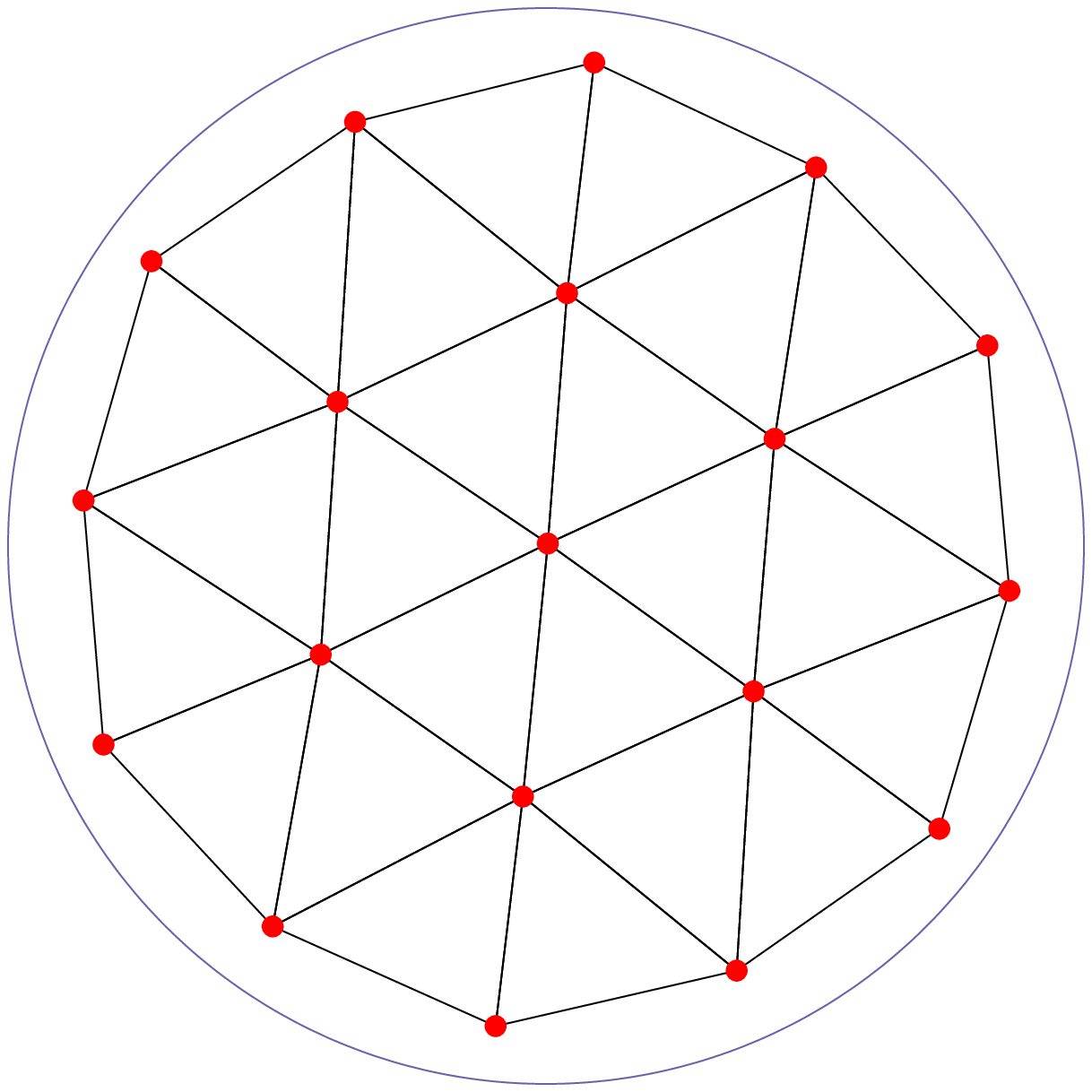}}
\hspace{-0.12in} 
\subfigure[]{
\includegraphics[width=0.66in]{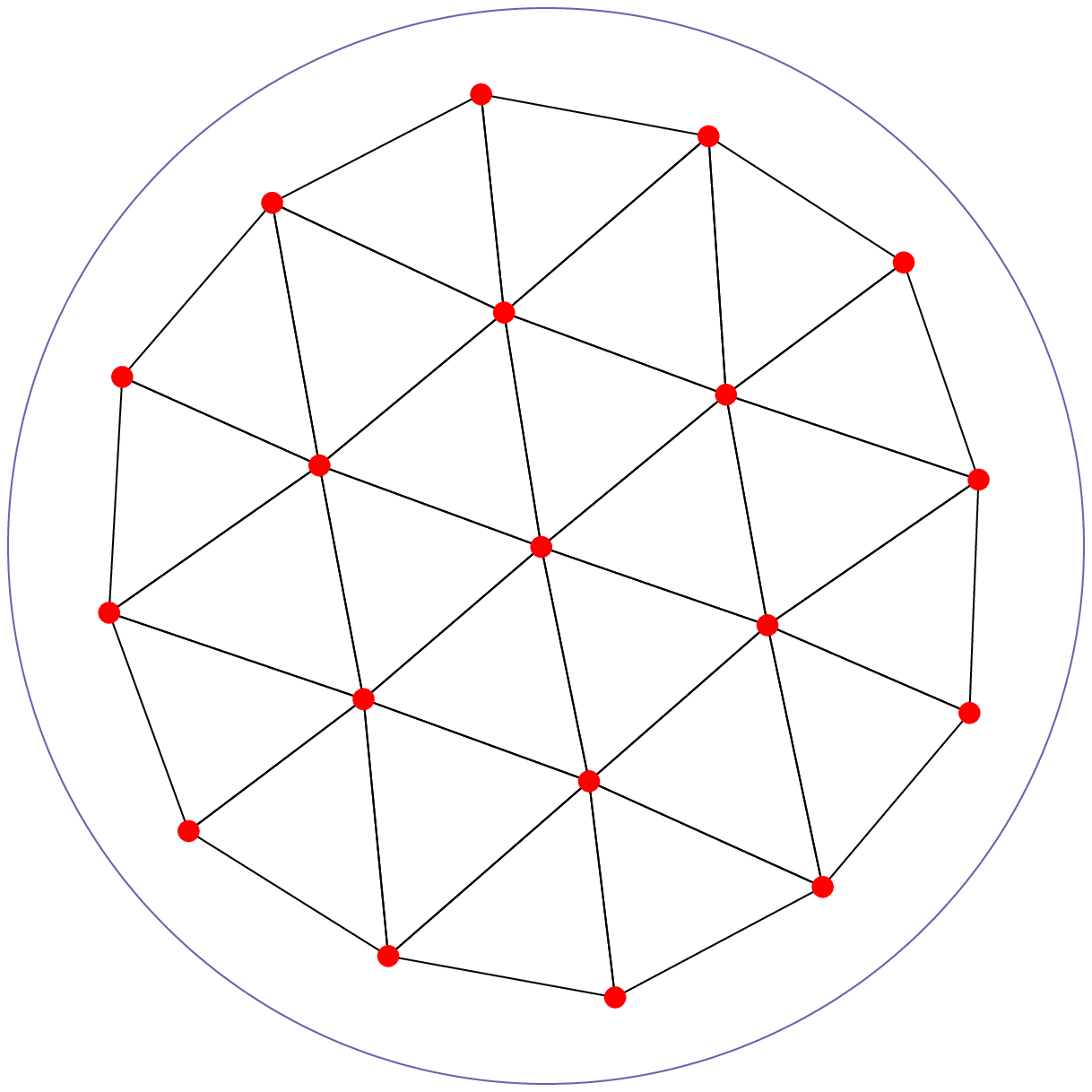}}
\hspace{-0.12in} 
\subfigure[]{
\includegraphics[width=0.66in]{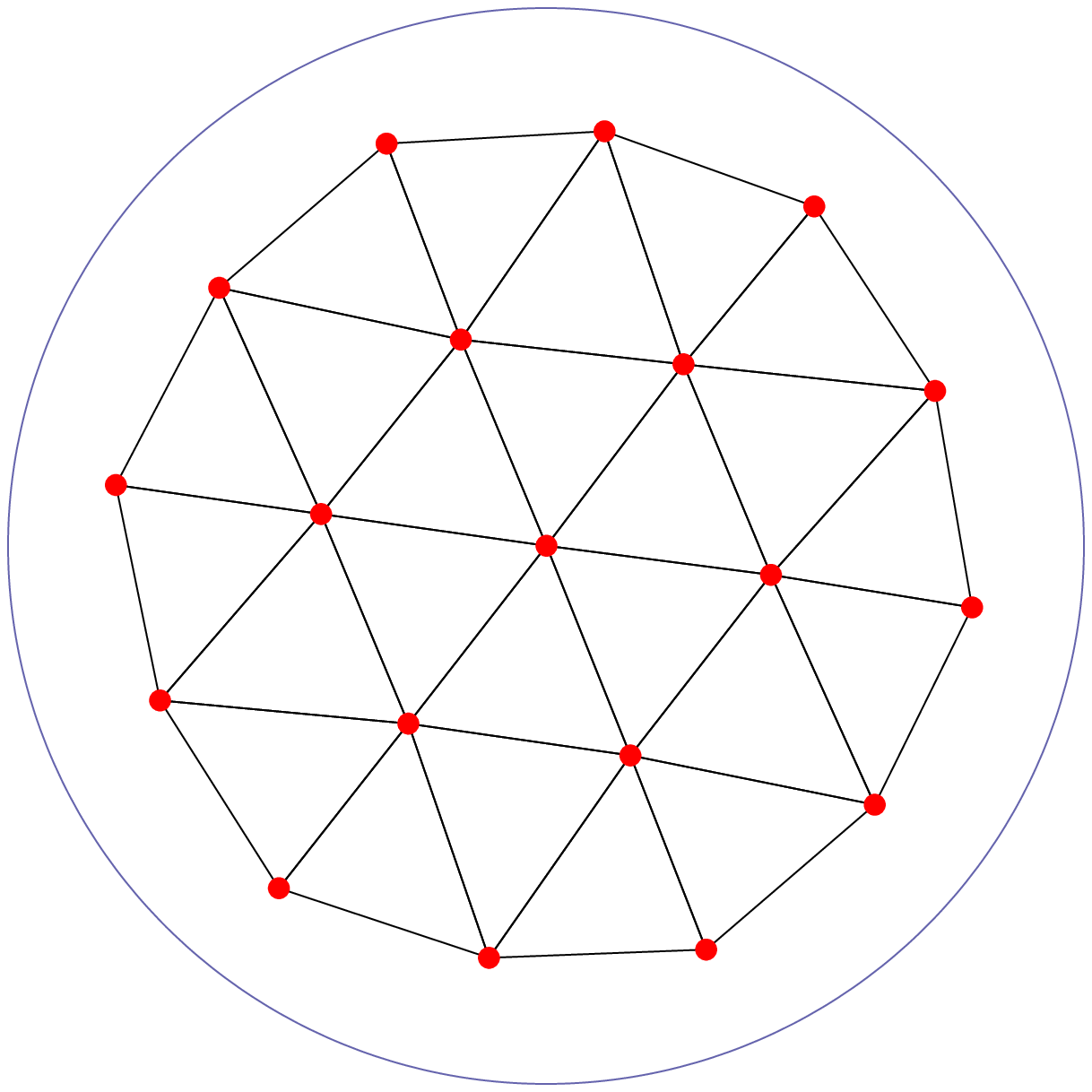}}
\caption{(Color online) The low-energy configuration of 19 filaments confined in a bundle. With
  the increase of $\Gamma$, the crystalline order emerges. The Delaunay
  triangulations are constructed in (c, e, f). $\Gamma = 2$ (a), $3$ (b), $4$ (c), $4.5$ (d), $5$ (e),
$6$ (f), $7$ (g), $8$ (h), $9$ (i), $10$ (j). $R=1$. }
\label{config_Gamma}
\end{figure}

In what follows, we will present typical results for small ($N=19$) and large
($N=50$) bundles. Figure~\ref{config_Omega} shows the low-energy configurations of 19 filaments
confined in a bundle subject to the confinement potential $V_{conf}$ with the
increase of the screening length (from left to right) that are generated via the
MC simulation. The comparison of the upper row ($\Gamma=0$) and the lower row
($\Gamma=1$) indicates that the confinement potential significantly facilitates
the formation of crystalline order; a hexagonal crystalline order has been well
established at $\kappa R=25$ for $\Gamma=1$, as shown in
Fig.~\ref{config_Omega}(e). In the regime of short screening length ($\kappa R
\gg 1$), since the confinement potential decays exponentially away from
the wall, the particles can only feel a strong repulsion from the wall if they
are within about one screening length from it. On the other hand, the
particles at a distance exceeding $\kappa^{-1}$ are invisible to one other.
Therefore, the system is essentially composed of $N$ soft
disks of effective radius $\kappa^{-1}$ confined in a disk of effective radius
$R-\kappa^{-1}$. With the increase of the screening length, the available area a particle
can explore is consequently reduced, and either a crystalline order or a glass state will
finally be formed at some critical value for the screening length. This scenario
is substantiated in the simulation. Fig.~\ref{config_Omega} (d-f) shows that the hexagonal
crystalline order starts to appear only if the
screening length exceeds some critical value $\kappa^{-1}  = 0.04$ as read from
the red curve in
Fig.~\ref{phi_Gamma}(a), which is corresponding to $4\ \textrm{nm}$ for the
typical value of $R=100\ \textrm{nm}$~\cite{cui2010spontaneous}.

\begin{figure}[t]  
 \centering 
\includegraphics[width=3in]{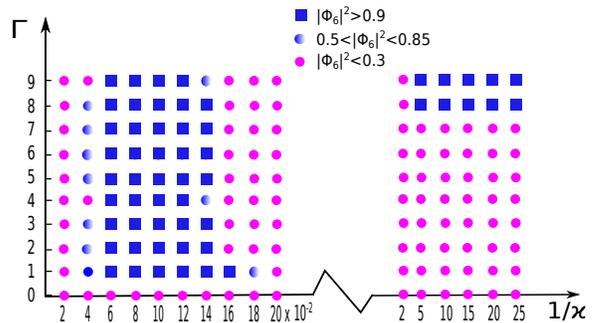}
\caption{(Color online) The phase diagram of filaments in a bundle in terms of the screening
  length $\kappa^{-1}$ and the phenomenological parameter $\Gamma$. The blue
  squares represent the
crystalline zone and the red dots are disordered states. $N=19$. $R=1$. }
\label{phase_diagram}
\end{figure}

\begin{figure}[t]  
\centering 
\subfigure[]{
\includegraphics[width=0.81in]{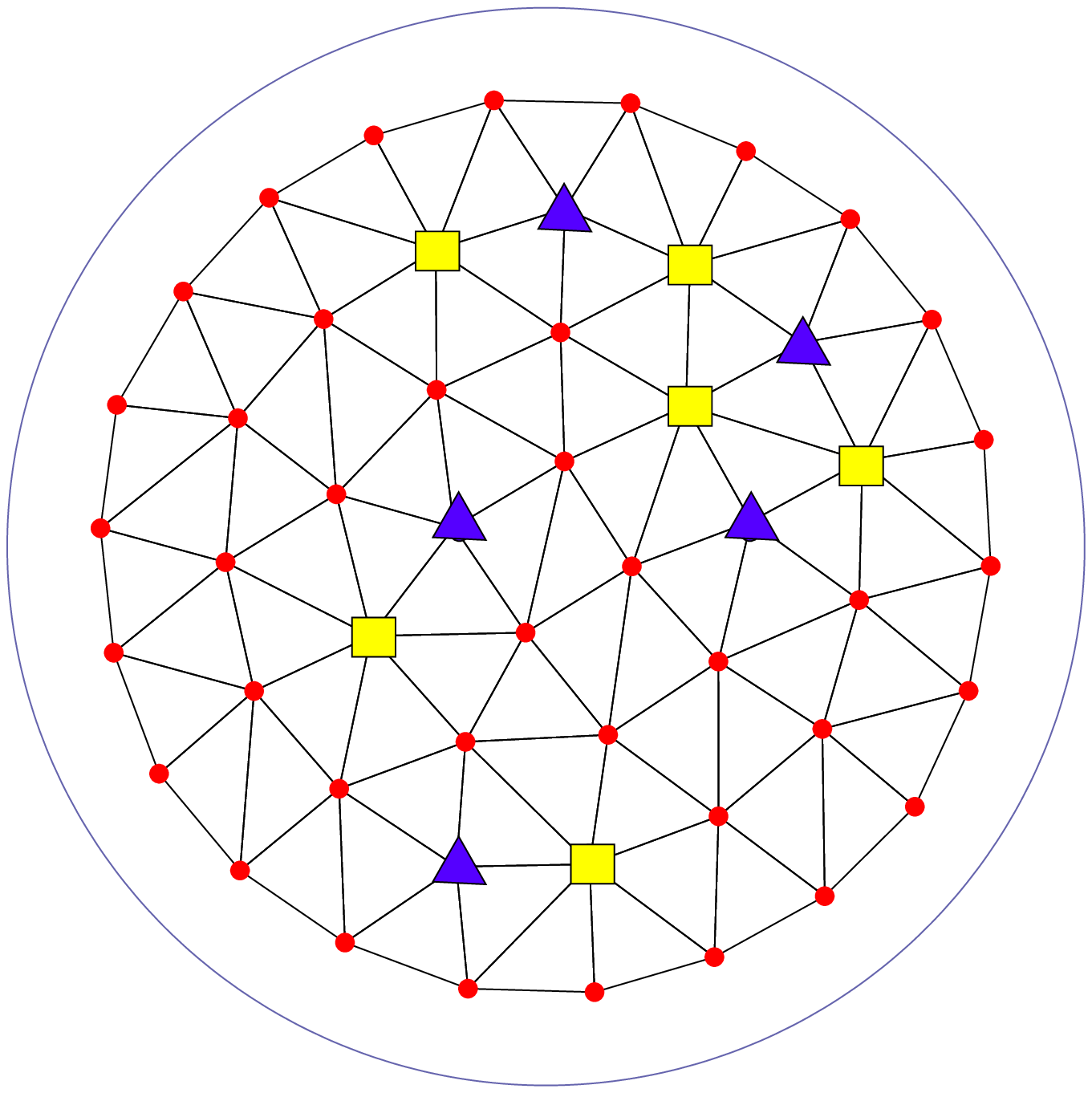}}
\hspace{0.1in} 
\subfigure[]{
\includegraphics[width=0.8in]{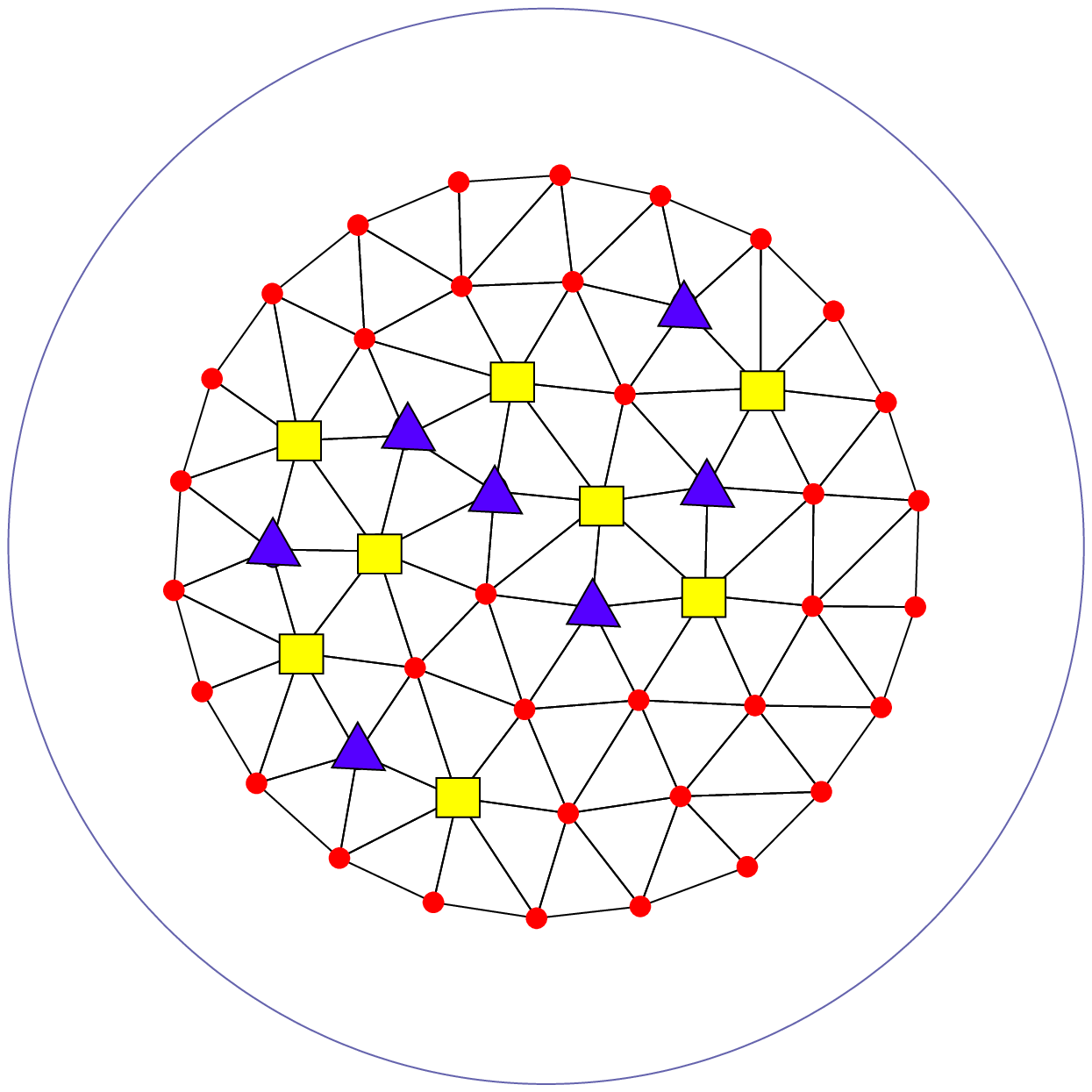}}
\caption{(Color online) The Delaunay triangulation of the low-energy configurations of 50 filaments.
    $\Gamma=1$ (a) and $10$ (b). $\kappa^{-1}
  =0.1$.
   The five-fold and seven-fold disclinations are
represented by blue triangles and yellow squares, respectively. $R=1$. }
\label{config_large_N}
\end{figure}

\begin{center}
\begin{table*}[t]
\caption{The distribution of the order parameter $|\Phi_6|^2$ for a bundle of
  $N=50$ filaments in the parameter space of $\Gamma$ and $1/\kappa$. $R=1$.
 } 
\centering
\begin{tabular}{|c|c|c|c|c|c|c|c|c|c|c|c|c|c|c|c|c|}
  \hline
  \backslashbox {$\Gamma$} {$1/\kappa$} & 0.02 & 0.04 & 0.06 & 0.08 & 0.1 & 0.12
																	  & 0.14 &0.16 & 0.18 & 0.2 & 2 & 5 & 10 & 15
													& 20 & 25 \\
  \hline
  9 & 0.00 & 0.12& 0.13& 0.15& 0.08& 0.36& 0.37& 0.36& 0.36& 0.35&
 0.02& 0.02& 0.00& 0.01& 0.08& 0.09\\
  \hline
  8 &  0.04& 
0.09& 
0.12& 
0.06& 
0.09& 
0.37& 
0.39& 
0.37& 
0.37& 
0.33& 
0.00& 
0.00& 
0.05& 
0.00& 
0.05& 
0.06\\
\hline
 7 &  0.00 & 
0.23& 
0.03& 
0.11& 
0.07& 
0.39& 
0.39& 
0.37& 
0.37& 
0.39& 
0.09& 
0.00& 
0.14& 
0.00& 
0.00& 
0.14\\
   \hline
  6 & 
0.00& 
0.13& 
0.04& 
0.12& 
0.11& 
0.05& 
0.38& 
0.37& 
0.39& 
0.36& 
0.16& 
0.00& 
0.01& 
0.00& 
0.00& 
0.00\\
  \hline
  5 &
0.00&
0.08&
0.03&
0.14&
0.11&
0.41&
0.39&
0.37&
0.37&
0.36&
0.00&
0.00&
0.00&
0.00&
0.00&
0.00\\ 
  \hline
  4 & 
0.02&
0.12&
0.14&
0.09&
0.40&
0.07&
0.38&
0.39&
0.38&
0.30&
0.00&
0.00&
0.00&
0.00&
0.00&
0.00\\
  \hline
  3 & 
0.02&
0.12&
0.34&
0.19&
0.11&
0.41&
0.37&
0.39&
0.36&
0.36&
0.01&
0.00&
0.00&
0.00&
0.00&
0.00\\
  \hline
  2 & 
0.00& 
0.22& 
0.09& 
0.11& 
0.11& 
0.39& 
0.37& 
0.39& 
0.37& 
0.38& 
0.00& 
0.00& 
0.00& 
0.00& 
0.00& 
0.00\\
\hline
1 & 
0.01&
0.13&
0.15&
0.11&
0.09&
0.07&
0.11&
0.38&
0.382&
0.40&
0.00&
0.00&
0.02&
0.00&
0.00&
0.01\\
  \hline
0 & 
0.00& 
0.04& 
0.06& 
0.26& 
0.31& 
0.15& 
0.14& 
0.24& 
0.06& 
0.05& 
0.00& 
0.00& 
0.01& 
0.03& 
0.00& 
0.01\\
  \hline
\end{tabular}
\label{table:ptei}
\end{table*}
\end{center}

We proceed to discuss the crystallization mechanism in the large screening
length limit, where the filaments in a bundle behave like a 2D
Coulomb gas in a disk, and the confinement potential conforms to a square law.
Without considering the confinement potential, the particles in the
zero-temperature 2D Coulomb gas are always uniformly distributed
along the circumference of the disk~\cite{sancho2001distribution}. This
remarkable feature
is specific to the logarithm potential. A confinement potential is therefore
required for pushing particles away from the boundary and forming some ordered
structure in the interior of the disk. 
Figure~\ref{config_Gamma} shows the low-energy configurations of 19 filaments
confined in a bundle with the increasing $\Gamma$ from $2$ (a) to $10$ (j),
where two transitions are identified. The first one occurs at
$\Gamma = 2$ where a particle is pushed from the boundary to the center of the
disk. In this jump, the reduction of the confinement potential exceeds
the energy barrier by moving a particle from the boundary to the center of the
disk. With the further increase of $\Gamma$, more and more
particles are pushed to the interior of the disk, forming a series
of symmetric patterns, as shown in Fig.~\ref{config_Gamma}(b-f). These discrete
structures break rotational symmetry, despite the existing rotational symmetry
in the
potential. As the total number of particles in the interior of the
disk exceeds six, the hexagonal crystalline structure emerges that is
highlighted by the Delaunay triangulation. The effect of further increase of $\Gamma$ is to
compress the system; the particles originally on the boundary start to migrate
towards the interior of the disk [see
Fig.~\ref{config_Gamma}(g-j)]. Fig.~\ref{phi_Gamma}(b) shows a rather
sharp disorder-order phase transition at $\Gamma=7$ that corresponds to the
configuration in Fig.~\ref{config_Gamma}(g).

Figure~\ref{phase_diagram} shows the phase diagram of the system in the
parameter space of $\kappa^{-1}$ and $\Gamma$. The two crystalline zones are
represented by blue squares. The interesting re-entrance effect for $\Gamma \ge 8$
is found in simulation. This agrees with general observations that confinement
effects yield re-entrance~\cite{bubeck1999melting,messina2003reentrant,royall2006re}. The
formation of the left crystalline zone in Fig.~\ref{phase_diagram} is understood
in terms of the soft-disk picture, while the upper right one is attributed to
the confinement potential that pushes particles away from the disk boundary, as
has been
discussed in the proceeding paragraphs. As
$\kappa^{-1}$ exceeds some critical value (about $0.14$ for $\Gamma \in [2,9]$
and $0.18$ for $\Gamma=1$), the crystalline order is destroyed. These critical
values are very close to half of the lattice spacings in the corresponding
crystallized filaments at $\kappa^{-1}=0.14$ for $\Gamma \in [2,9]$ ($d=0.33$)
and at $\kappa^{-1}=0.18$ for $\Gamma =1$ ($d=0.39$), respectively. The melting
of the crystals is therefore driven by increasing the effective radius of the
soft disks; the melting starts when the repulsion between particles becomes
strong enough so that the confinement potential fails to hold the particles
together. The simulation also indicates that the crystalline order can be
destroyed for $\Gamma$ exceeding about $50$ and $100$ in the short- and
large-screening length regimes, respectively. The underlying physics is the
overcompression-induced breakage of a crystal; the compression originates from
the confinement potential tends to push particles towards the center of the
disk. 

As the number of particles increases, the value of the order parameter
$|\Phi_6|^2$ is generally reduced, as shown in Table~\ref{table:ptei} for a
bundle of $50$ filaments in the parameter space of $\Gamma$ and $1/\kappa$. The
maximum value for the order parameter in the region considered in
Table~\ref{table:ptei} does not exceed 0.5, and the value of the order parameter
for large screening length is even lower. It implies that a large system tends
to be in a disordered state.  The emerging topological defects in large systems
are responsible for the reduction of the value of the order parameter; their
proliferation destroys the crystalline order. Figure~\ref{config_large_N} shows
the five- and seven-fold disclinations in a bundle of $50$ filaments. It is
important to note that these defects are introduced via physical potentials
instead of either a non-Euclidean background geometry~\cite{bowick2009two} or
geometrically induced stresses~\cite{PhysRevLett.105.045502}.  In simulation, we
take attempts to reduce the possibility of artificially introducing defects,
such as choosing various initial configurations and carefully heating the
lowest-energy states repeatedly to avoid the metastable states. The
irremovability of the topological defects in simulation makes one to conjecture
that defects may exist intrinsically in large bundles that are subject to a
spatially varying potential.

\section{Conclusion}

Our particle model shows that the interplay between the repulsive interaction
and network confinement leads to the re-entrance phenomenon in the phase
diagram. In addition, MC simulation suggests the emergence of topological
defects in large bundles via pure physical potentials. This may lead to further
study about the formation mechanism of topological defects in two-dimensional
systems. Our model provides an example of controlling the separation of
filaments and their bundling that may find applications in the control of cells
in external filamentous matrices~\cite{meredith1993extracellular} and the design
of biomaterials. In addition, the electrostatic repulsion-driven crystallization
model arising from the study of filament networks can even find a more general
context. Crystallization, melting and dynamics of confined two-dimensional
charged colloidal systems have been extensively studied, where the particles are
mutually repelled~\cite{assoud2008binary, lin_dust,lowen2012colloidal}. In our
model, the introduced spatially varying confinement potential that is mimicking
the charged environment of a bundle can find its applications in a general
colloidal system.

\section*{ACKNOWLEDGMENTS}

This work was funded by grants from the Office of the Director of
Defense Research and Engineering (DDR$\&$E) and the Air Force
Office of Scientific Research (AFOSR) under Award No.
FA9550-10-1-0167.

\bibliography{filament_ref}
\end{document}